\documentclass[12pt]{article}
\usepackage{amsmath,amssymb,epsfig,color, url}
\usepackage{psfrag}
\usepackage{graphicx}
\usepackage{bm}

\newcommand{\be}{\begin{equation}}  
\newcommand{\ee}{\end{equation}}

\begin{document}

\begin{titlepage}

\begin{flushright}
WSU-HEP-1401\\
July 21, 2014
\end{flushright}

\vspace{0.7cm}
\begin{center}
\Large\bf 
Model independent extraction of 
the proton magnetic radius from electron scattering
\end{center}

\vspace{0.8cm}
\begin{center}
{\sc  Zachary Epstein$^{(a)}$,  Gil Paz$^{(b)}$, Joydeep Roy$^{(b)}$}\\
\vspace{0.4cm}
{\it $^{(a)}$ Department of Physics, University of Maryland, College Park, Maryland 20742, USA
}\\
\vspace{0.3cm}
{\it 
$^{(b)}$ 
Department of Physics and Astronomy \\
Wayne State University, Detroit, Michigan 48201, USA 
}

\end{center}
\vspace{1.0cm}
\begin{abstract}
  \vspace{0.2cm}
  \noindent
We combine constraints from analyticity with experimental 
electron-proton scattering data to determine the proton magnetic 
radius without model-dependent assumptions on the shape of the form factor. 
We also study the impact of including electron-neutron 
scattering data, and $\pi\pi\to N\bar{N}$ data.
Using representative datasets we find for a cut of $Q^2\leq0.5$ GeV$^2$,
$r_M^p=0.91^{+0.03}_{-0.06}\pm0.02$ fm using just proton scattering data;
$r_M^p=0.87^{+0.04}_{-0.05}\pm0.01$ fm adding neutron data; and 
$r_M^p=0.87^{+0.02}_{-0.02}$ fm adding $\pi\pi$ data. We also extract the neutron magnetic radius from these data sets obtaining $r_M^n=0.89^{+0.03}_{-0.03}$ fm from the combined proton, neutron, and $\pi\pi$ data.
\end{abstract}
\vfil

\end{titlepage}

\section{Introduction}
The first indication of the composite nature of the proton was the measurement of the magnetic moment of the proton by Frisch and Stern in 1933 \cite{Frisch}. As described by Otto Stern in his Nobel prize lecture, ``The result of our measurement was very interesting. The magnetic moment of the proton turned out to be about 2.5  times larger than the theory predicted. Since the proton is a fundamental particle - all nuclei are built up from protons and neutrons - this result is of great importance. Up to now the theory is not able to explain the result quantitatively." \cite{Stern}. This statement is to some extent still true today.  The response of the proton to electromagnetic field is described by two form factors, one ``electric" ($G_E$) and one ``magnetic" ($G_M$). The magnetic moment of the proton is just the value of $G_M$ at zero 4-momentum transfer squared. Viewed as a Taylor series, the magnetic moment is the first in an infinite list of numbers needed to describe the response of the proton to a magnetic field. The next number would be the slope of the magnetic form factor at zero, which is related to the magnetic radius of the proton. For the electric form factor the value at zero is the total charge of the proton in units of $e$, and the slope at zero defines the charge radius of the proton.  The electric and magnetic radii of the proton are therefore as fundamental as the charge and magnetic moment of the proton. Currently, we cannot determine them accurately from theory, although lattice QCD is making progress on this issue; see for example \cite{Green:2014xba}. We can measure them from experiment. 

The determination of the charge radius of the proton has received considerable attention in the last few years as a result of the discrepancy between the extraction of the charge radius of the proton from muonic and regular hydrogen. The measurement reported by the CREMA collaboration in \cite{Pohl:2010zz} has found $r_E^p=0.84184(67)$~fm, and more recently \cite{Antognini:1900ns} $r_E^p=0.84087(39)$~fm. Both of these muonic hydrogen extractions are in conflict with the CODATA 2010 \cite{Mohr:2012tt} value $r_E^p=0.87580(770)$~fm, based on only hydrogen and deuterium spectroscopic data. This discrepancy is often referred to as the ``proton radius puzzle." 

The discrepancy has generated considerable debate.  The discussion has focused on the one hand on recalculation of the theoretical input to the extraction of $r_E^p$ from muonic hydrogen and on modifications of the theoretical calculation such as proton structure effects, e.g. \cite{Daza:2010rh, Jentschura:2010ej, Hill:2011wy, Carroll:2011rv, Eides:2012ue, Kelkar:2012hf, Borie:2012zz, Antognini:2012ofa, Indelicato:2012pfa, Graczyk:2013pca, Chen:2013udl, Giannini:2013bra, Korzinin:2013uia, Indelicato:2014mra, Faustov:2014pwa, Karshenboim:2014maa, Friedmann:2009mx, Friedmann:2009mz,  DeRujula:2010dp, Cloet:2010qa, Vanderhaeghen:2010nd,  DeRujula:2010ub, Kholmetskii:2010sx,DeRujula:2010zk, Distler:2010zq, Miller:2011yw, Carlson:2011zd,  Pineda:2011xp,  Wu:2011jm, Carlson:2011dz, Kholmetsky:2012zz, Karr:2012mfa, Birse:2012eb, Miller:2012ht, Miller:2012ne, Greenberg:2012vn, Gorchtein:2013yga, Mart:2013gfa, Mohr:2013axa, Robson:2013nwa, Alarcon:2013cba, Downie:2013fya, Jentschura:2014ila, Eides:2014swa, Peset:2014yha, Gainutdinov:2014kma, Tomalak:2014dja, Pachucki:2014zea, Glazek:2014ria, Gorchtein:2014hla, Peset:2014jxa}, and on effects of new physics, e.g. \cite{Jaeckel:2010xx, Kauffmann:2010cu, Brax:2010gp, Barger:2010aj,TuckerSmith:2010ra, Batell:2011qq, Rivas:2011dm, Barger:2011mt, Carlson:2012pc, Wang:2013fma, Li:2013dwa, Moumni:2013yta, Carlson:2013mya, Chang:2013yva, Onofrio:2013fea, Karshenboim:2014tka, Ubachs:2013kpa, Brax:2014zba} on the other. 

Apart from regular and muonic hydrogen, electron proton scattering data also allows to measure the charge radius of the proton. Many such extractions exist in the literature, using different data sets and functional forms. The main problem in robust extraction of the proton charge radius from the data is the need to reliably extrapolate the form factor to $q^2=0$ in order to find its slope. Many of the existing  extractions postulate a functional form for the form factor either explicitly, or implicitly by truncating a possibly general series expansion. Thus all of these extractions introduce model dependance for the value of $r_E^p$ which is very hard to assess. 

\begin{figure*}
\centering
\includegraphics[height=0.2\textwidth]{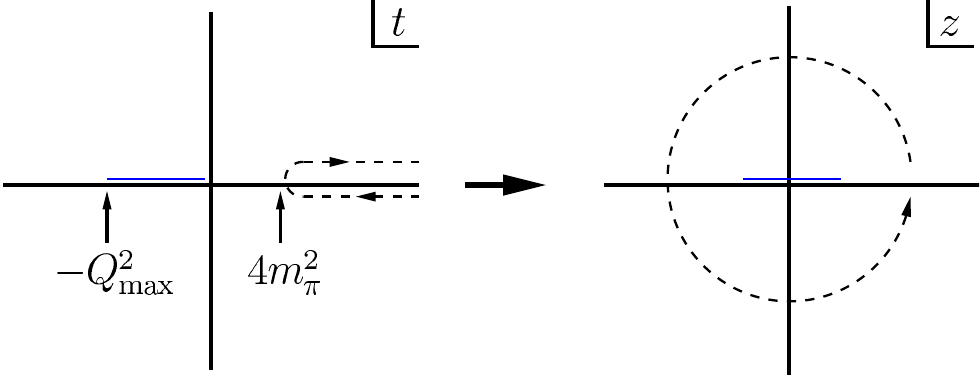}
\caption{ \label{cutplane} Conformal mapping of the cut plane to the unit circle.}
\end{figure*}

\begin{figure}[h!]
\begin{center}
\includegraphics[scale=0.65]{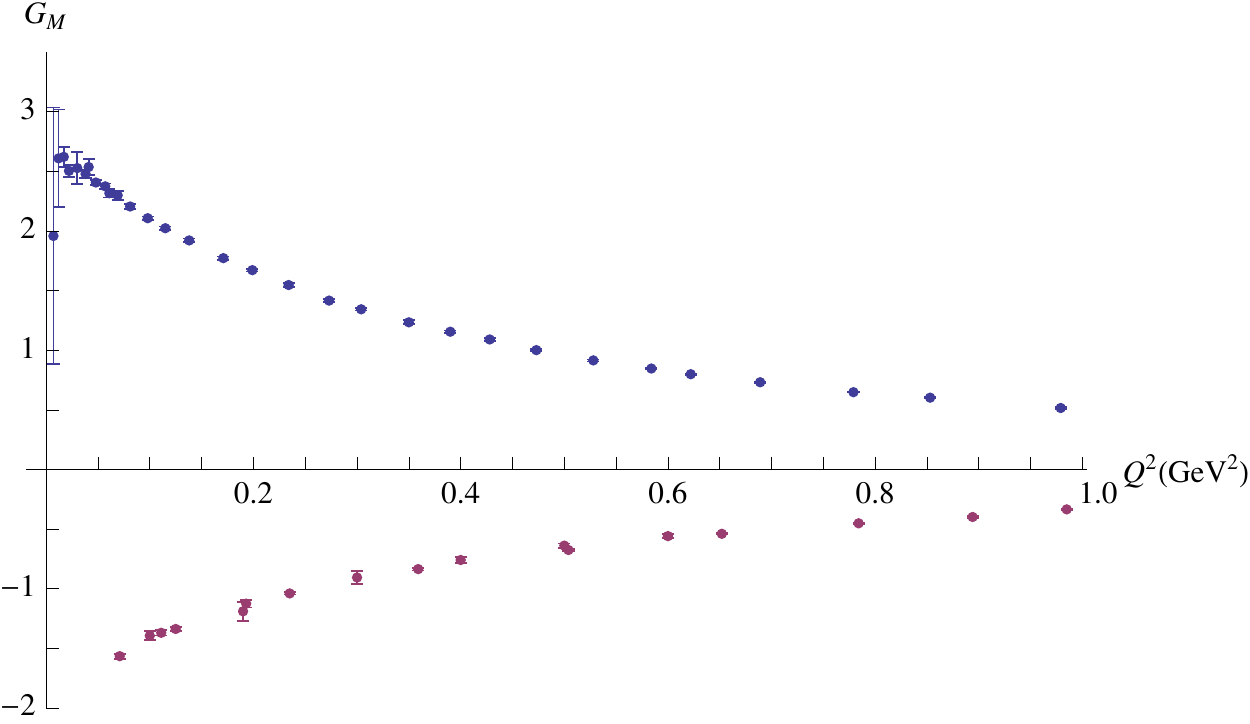}\quad
\includegraphics[scale=0.6]{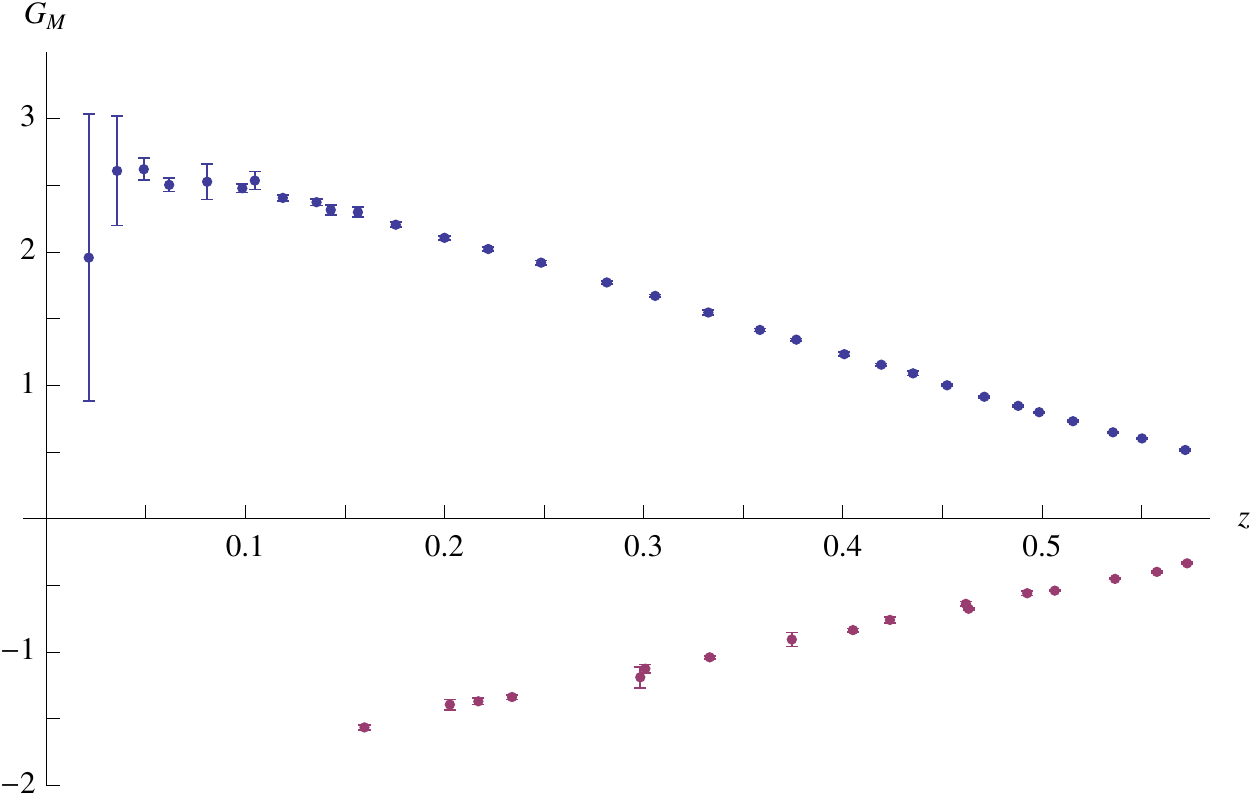}
\caption{\label{GM} Proton (above the horizontal axes) and neutron (below the horizontal axes) magnetic form factor data as a function of $Q^2$ (left) and as a function of $z$ (right).  
Here we choose $t_0=0$ and use $t_{\rm cut}=4m_\pi^2$ in the definition of $z$, 
and plot data 
for $0\le Q^2 \le 1.0\,{\rm GeV}^2$.   
}
\end{center}
\end{figure}

The problem was solved in  \cite{Hill:2010yb}, which introduced a method of extraction that is free of such model dependance. The method, often called the ``$z$ expansion" adapts an established tool in the study of \emph{meson} form factors to the case of \emph{baryon} form factors. The $z$ expansion relies on the known analytic properties of the electromagnetic form factors $G_E$ and $G_M$. They are analytic in the complex plane outside of a cut along the positive real $q^2$ axis that starts at $4m_\pi^2$ and extends to infinity. The location of the singularity also implies that the radius of convergence, if using a simple Taylor expansion for the form factors, is at  most  $4m_\pi^2$. Most of the data about the form factors is well above this value. But even if we use data that is strictly below it, it is questionable whether we can ignore higher terms in the Taylor expansion as it is often assumed. The $z$ expansion avoids this difficulty. By using the variable $z$ defined as 
\begin{equation}\label{z}
z(t,t_{\rm cut},t_0) = {\sqrt{t_{\rm cut} - t} - \sqrt{t_{\rm cut} - t_0} \over \sqrt{t_{\rm cut} - t} + \sqrt{t_{\rm cut} - t_0}  }
\end{equation}
we can map the domain of the analyticity of the form factors onto the unit circle; see Figure~\ref{cutplane}. 
For  $G_E$ and $G_M$, $t_{\rm cut}=4m_\pi^2$. The free parameter $t_0$ determines the location of $z=0$. Considered as  a function of $z$, the form factor is analytic inside the unit circle and can be expressed as
\begin{equation}
G_{E,M}(q^2)=\sum_{k=0}^\infty a_k\, z(q^2)^k.
\end{equation} 
Intuitively, $z$ is the ``right" variable in which to perform a Taylor expansion of the form factor. Unlike a Taylor expansion in $q^2$,  the expansion is guaranteed to converge for $|z|<1$. Since for finite negative $q^2$, $z$ is smaller than 1, this guarantees convergence for any $q^2$ measured in experiment.  As an illustration to this intuitive picture, consider the proton magnetic form factor data  tabulated in \cite{Arrington:2007ux} and the neutron magnetic form factor data  tabulated in \cite{Lung:1992bu,Gao:1994ud, Anklin:1994ae,Anklin:1998ae, Kubon:2001rj, Anderson:2006jp, Lachniet:2008qf}. Plotting the data points as a function of $Q^2=-q^2$ for $0<Q^2\leq 1$ GeV$^2$,  we see a considerable curvature; see Figure \ref{GM}.  If we plot the same data as a function of $z$ (using $t_{\rm cut}=4m_\pi^2$ and $t_0=0$) the data looks fairly linear. We can also easily estimate the slopes of the proton and neutron magnetic form factors. If we plot the normalized values of the form factors, i.e. the form factor values divided by their value at $q^2=0$ as a function of $z$, the slopes would be hard to distinguish. This implies that the magnetic radii of the proton and neutron are very similar. We will see later that this is indeed the case.  

The magnetic radius of the proton is defined as $r_M^p \equiv \sqrt{\langle r^2\rangle_M^p}$, where   
\begin{equation}\label{definition}
\langle r^2 \rangle_M^p=
\frac{6}{G_M^p(0)}\frac{d}{dq^2}G_M^p(q^2)\bigg|_{q^2=0}\, .
\end{equation}

In 2010 the A1 collaboration reported a value of  $r_M^p=0.777(13)_{\rm stat.}(9)_{\rm syst.}(5)_{\rm model}(2)_{\rm group}$ fm \cite{Bernauer:2010wm}. This value is considerably lower than $r_M^p=0.876\pm0.010\pm0.016$ fm extracted in \cite{Borisyuk:2009mg} or $0.854 \pm 0.005$ fm extracted in \cite{Belushkin:2006qa}, the  two other extractions cited by the Particle Data Group (PDG)  \cite{Beringer:1900zz}. Are we facing also a \emph{magnetic} radius puzzle?\footnote{See the conclusions for values of $r_M^p$ not quoted by the PDG.} 

The purpose of this study is to apply the methods established in \cite{Hill:2010yb}, to the extraction of the magnetic radius of the proton from scattering data.  As in \cite{Hill:2010yb} we will use proton, neutron, and $\pi\pi$ scattering data to determine the magnetic radius of the proton from the reported measurement of the magnetic form factors of the proton and  neutron.  We will also determine the magnetic radius of the neutron.

The rest of the paper is organized as follows. In section \ref{constraints} we discuss the analytic structure of the form factors and their constraints. In section \ref{proton} we extract the magnetic radius of the proton from proton, neutron, and $\pi\pi$ scattering data. In section \ref{neutron}  we extract the magnetic radius of the neutron from the same data. We present our conclusions in section \ref{conclusions}.

\section{Form factor constraints}\label{constraints}
The analytic structure of the form factors and their constraints were discussed in detail in \cite{Hill:2010yb}. Here we review some of the main ingredients needed for our analysis.
\subsection {Form factor definitions}
The Dirac and Pauli form factors, $F^N_1$ and $F^N_2$, 
respectively, are defined by \cite{Foldy:1952,Salzman:1955zz}
\be\label{DP}
\langle N(p')|J_\mu^{\rm em}|N(p)\rangle=\bar u(p')
\left[\gamma_\mu F^N_1(q^2)+\frac{i\sigma_{\mu\nu}}{2m_N}F^N_2(q^2)q^\nu\right]u(p)\,,
\ee
where $q^2=(p'-p)^2=t$ and $N$ stands for $p$ or $n$. 
The Sachs electric and magnetic form factors
are related to the Dirac-Pauli basis by \cite{Ernst:1960zza}
\begin{align}
G^N_E(t) = F^N_1(t) + {t\over 4m_N^2} F^N_2(t) \,, \quad
G^N_M(t) = F^N_1(t) + F^N_2(t) \,. 
\end{align}
At $t=0$ they are~\cite{Beringer:1900zz}
$G^p_E(0) = 1$,
$G^n_E(0) = 0$,
$G^p_M(0) = \mu_p \approx 2.793$, 
$G^n_M(0) = \mu_n\approx -1.913$.
We define the isoscalar and isovector form factors as
\begin{align}\label{iso}
G_{M,E}^{(0)} = G_{M,E}^{p} + G_{M,E}^{n} \,, \quad
G_{M,E}^{(1)} = G_{M,E}^{p} - G_{M,E}^{n} \,, 
\end{align} 
such that at $t=0$ they are
$G_E^{(0)}(0) = 1$, 
$G_E^{(1)}(0) = 1$,  
$G_M^{(0)}(0) = \mu_p+\mu_n$, 
$G_M^{(1)}(0) = \mu_p-\mu_n$.
Notice that $G^{(0)}_{M,E}=2G^{s}_{M,E}$, $G^{(1)}_{M,E}=2G^{v}_{M,E}$ for $G^{s,v}_{M,E}$ of \cite{Belushkin:2006qa}.
\subsection{Analytic structure}
The electric and magnetic form factors are analytic functions of $t$ outside of a cut that starts at the two-pion threshold $t\geq 4m_\pi^2$ on the real $t$ axis. The scattering data lies on $-Q^2_{\rm max}\leq t\leq0$, where $Q^2_{\rm max}$ denotes the largest value of $Q^2$ in a given data set. The domain of analyticity can be mapped onto the unit disk via the conformal transformation (\ref{z}). The mapping is shown in Figure \ref{cutplane}. The maximal value of $|z|$ depends on $Q^2_{\rm max}$ and $t_0$. It is minimized for the choice 
$t_0^{\rm opt} = t_{\rm cut} \left( 1 - \sqrt{1+ Q^2_{\rm max}/t_{\rm cut}} \right)$ which is also the value used for Figure \ref{cutplane} .

Since the values of the form factors at $q^2=0$ are well known, in the following we will use $t_0=0$. As discussed in \cite{Hill:2010yb}, the results do not depend on the choice of $t_0$. For this choice of $t_0$, the maximum value of $|z|$ is $0.46, 0.58$ for $Q^2_{\rm max}=0.5,1.0$ GeV$^2$, respectively . The form factors can be expanded in a  power series in $z(q^2)$:
\begin{equation} 
G(q^2) = \sum_{k=0}^\infty a_k \, z(q^2)^k \,,
\end{equation}
where higher order terms are suppressed by powers of the maximum values of $|z|$.  The coefficients $a_k$ are also bounded in size guaranteeing that the series converges.  

The analytic structure implies the dispersion relation
\begin{align}\label{dispersion}
G(t) = {1\over \pi} \int_{t_{\rm cut}}^\infty {dt^\prime}\, {{\rm Im}G(t^\prime + i0) \over t^\prime - t } \,. 
\end{align} 
Information about ${\rm Im}G$ over the cut can be translated into information about $a_k$.  As shown in  \cite{Hill:2010yb}, we have 
\begin{align}\label{Fourier} 
a_0&=\frac{1}{\pi}\int_0^\pi d\theta\,{\rm Re}\, G[ t(\theta)+i0]=G(t_0) \,,
\nonumber\\
a_k&=-\frac{2}{\pi}\int_0^\pi d\theta\,  {\rm Im}\, G[t(\theta)+i0 ] \,\sin(k\theta) 
= {2\over \pi} \int_{t_{\rm cut}}^\infty {dt\over t-t_0} \sqrt{ t_{\rm cut} - t_0 \over t - t_{\rm cut}} 
{\rm Im}G(t)
\sin[ k\theta(t) ] 
\,,\quad k\ge 1\,,
\end{align}
where
\begin{equation}
t=t_0+\frac{2(t_{\rm cut}-t_0)}{1-\cos \theta}\equiv t(\theta)\,. 
\end{equation}
\subsection{Bounds on the coefficients}
In order to obtain a reliable and conservative extraction of the proton magnetic radius we need to establish appropriate bounds on the coefficients $a_k$. In particular, it was shown in  \cite{Hill:2010yb} that the bounds of $|a_k|<5$ and $|a_k|<10$ are very conservative for the electric form factor.  We would like to determine similar bounds for the magnetic form factor. 
\subsubsection{Vector dominance ansatz}

The first approach we use to estimate the size of $a_k$ is the vector dominance ansatz, where the  form factors are assumed to be dominated by  vector meson exchange: $\omega$ for $I=0$, and $\rho$ for $I=1$ \cite{Hohler:1974ht}.  In particular, the imaginary part of the form factor is given by \cite{Schwinger}
\be\label{VMD}
{\rm Im}G(t+i0) ={ {\cal N} m_{V}^3 \Gamma_{V}  \over (t-m_{V}^2)^2  + \Gamma_{V}^2 m_{V}^2 } \theta(t-t_{\rm cut})\,,
\ee
where $m_V$ and $\Gamma_V$ are the mass and width, respectively,  of the vector meson and ${\cal N}$ is a normalization constant determined below. Also, $t_{\rm cut}=9m_\pi^2$ for $I=0$ and $t_{\rm cut}=4m_\pi^2$ for $I=1$. 
Using the dispersion relation (\ref{dispersion}) with (\ref{VMD})
we find \cite{Bhattacharya:2011ah}
\begin{equation}
G(t+i0)=\frac{{\cal N}}\pi
\frac{m_{V}^3 \Gamma_{V}}
{|b(t)|^2}
\Bigg[\frac{1}{2}
\log\left(
\frac{|b(t_{\rm cut})|^2}{|t_{\rm cut}-t|^2}
\right)
+ { m_{V}^2 - t \over m_{V}\Gamma_{V}} \arg[b(t_{\rm cut})] 
+ i\pi \theta(t-t_{\rm cut})\Bigg]\,,
\end{equation}
where $b(t)=t-m_{V}^2+ i \Gamma_{V} m_{V}$, and  
${\cal N}$ is determined by the value of $G(0)$. 

This form allows us to calculate $a_k$ explicitly from (\ref{Fourier}). Using $m_\rho = 0.775$ GeV, 
$\Gamma_\rho = 0.149$ GeV, $m_\omega = 0.783$ GeV and $\Gamma_\omega = 0.0085$ GeV \cite{Beringer:1900zz}, we have  for $I=0$: $a_0\approx 0.88$, $a_1\approx 1.0$, $a_2\approx 0.83$, $a_3\approx -0.29$, $a_4\approx -1.1$. For $I=1$, we have $a_0 \approx4.7$, $a_1\approx 3.7$, $a_2\approx 2.7$, $a_3\approx 2.0$, $a_4\approx-0.36$. 
Also, using $|\sin(k\theta) |\leq 1$ allows us to obtain a $k$-independent bound on $a_{k}$ for $k\geq1$
\be
\left|{a_k \over a_0}\right| 
\le 
\frac{ 2 |{\cal N}|}{|G_M(t_0)|}  
{\rm Im} \left( { - m_{V}^2 \over b(t_{\rm cut}) + \sqrt{(t_{\rm cut}-t_0)b(t_{\rm cut})} } 
\right) \,.
\ee

 We find that $|a_k/a_0|\leq1.3$ for $I=0$ and $|a_k/a_0|\leq1.1$ for $I=1$. These results are very similar to those of  \cite{Hill:2010yb}. An important difference from the electric case is that  the magnetic form factors at $q^2=0$ are given by $G_{M}^{(0)}(0)\approx 0.88$ and $G_{M}^{(1)}(0)\approx 4.7$, compared to $G_{E}^{(0,1)}(0)=1$.  Since the vector dominance ansatz is normalized by the value at $q^2=0$, the coefficients $a_k$ are proportional to this value. Thus we find that $|a_k|\leq1.1$ for $I=0$ and $|a_k|\leq5.1$ for $I=1$.  We conclude that while $|a_k|\leq 10$ is a conservative estimate for this ansatz, a bound of $5$, namely $|a_k|\leq 5$ is not conservative enough.
\subsubsection{Explicit $\pi\pi$ continuum}\label{sec232}
For the case of the magnetic isovector form factor the singularities that are closest to the cut arise from the two-pion continuum. The imaginary part of  $G_M^{(1)}$ close to the cut can be described by the pion form factor $F_\pi(t)$ (normalized to $F_\pi(0)=1$) and $f^1_-(t)$, a partial $\pi\pi\to N\bar N$ amplitude \cite{Federbush:1958zz, Frazer:1960zzb, Belushkin:2006qa}:
\begin{align}\label{pipiNN}
{\rm Im}\, G_M^{(1)}(t) = \sqrt{\frac2{t}} \left(t/4-m_\pi^2 \right)^\frac32 F_\pi(t)^* f^{1}_-(t) \,.
\end{align}
Since $F_\pi(t)$ and $f^1_-(t)$ share the same phase \cite{Frazer:1960zzb}, we will replace them in (\ref{pipiNN}) by their absolute values \cite{Hill:2010yb}. The relation (\ref{pipiNN}) holds only up to the four-pion threshold  $t\le 16 m_\pi^2$, but in order to estimate the bounds on the coefficients, we will extend  (\ref{pipiNN}) through the $\rho$ peak as in \cite{Hill:2010yb}.\\
Values of $f_-^1(t)$ are taken from Table~2.4.6.1 of \cite{Hohler}. We interpolate their product with the prefactor in (\ref{pipiNN}). This interpolated function is multiplied by the values of  $|F_\pi(t)|$ using the four $t$ values from \cite{Amendolia:1983di}  ($0.101$ to $0.178\,{\rm GeV}^2$) and the 43 $t$ values from \cite{Achasov:2005rg} 
($0.185$ to $0.94\,{\rm GeV}^2$).  This gives us a discrete expression with 47 data points for ${\rm Im}\, G_M^{(1)}(t)$ from $0.101 {\rm GeV}^2$ to $0.94\,{\rm GeV}^2$. We now use the experimental data up to $t=0.8\,{\rm GeV}^2\approx 40\,m_\pi^2$ to calculate $a_k$ using (\ref{Fourier}). We find $a_0\approx 7.9$, $a_1\approx -5.5$, $a_2\approx -6.1$, $a_3\approx  -2.9$, $a_4\approx 1.1$. Using $|\sin(k\theta) |\leq 1$ gives $|a_k| \lesssim 7.2$ for $k\ge 1$. It is interesting to note that $a_k/a_0$ for these values is very similar to the analogous $a_k/a_0$ obtained for the isovector electric form factor in \cite{Hill:2010yb}. This can be traced to the fact that the shape of $ f^{1}_-(t)$ is very similar to $f^{1}_+(t)$. This indicates that the main difference between the electric and magnetic form factors is their normalization. 

\subsubsection{Bounds on the $t\geq4m_N^2$}
Above the two nucleon threshold one can use $e^+e^-\to N\bar N$ data to constrain the electric and magnetic form factor. In particular, the cross section is given by \cite{Cabibbo:1961sz}
\begin{align}
\sigma(t) = {4\pi\alpha^2\over 3 t} \sqrt{ 1 - {4m_N^2\over t}} \left( |G_M(t)|^2 + {2m_N^2\over  t} |G_E(t)|^2 \right) \,.
\end{align}
The contribution to $a_k$ from this region is given by (\ref{Fourier})
\begin{equation}
\delta a_k={2\over \pi} \int_{4m_N^2}^\infty {dt\over t-t_0} \sqrt{ t_{\rm cut} - t_0 \over t - t_{\rm cut}} 
{\rm Im}G(t)
\sin[ k\theta(t) ] 
\,,\quad k\ge 1\,.
\end{equation}
Since $|{\rm Im}G(t)\sin[ k\theta(t) ] |\leq|G_M(t)|\leq \sqrt{ |G_M(t)|^2 + \dfrac{2m_N^2}{t} |G_E(t)|^2}$ we have 
\begin{equation}
|\delta a_k|\leq{2\over \pi} \int_{4m_N^2}^\infty {dt\over t-t_0} \sqrt{ t_{\rm cut} - t_0 \over t - t_{\rm cut}} \,\sigma(t)\dfrac{3t}{4\pi\alpha^2}\left(1 - {4m_N^2\over t}\right)^{-1/2}\,,\quad 
 k\ge 1\,.
\end{equation}
These bounds are valid for both the proton and neutron magnetic form factors. 
Using the $e^+e^-\to p\bar p$ data from \cite{Ablikim:2005nn}, we perform the integral from $t=4.0\,{\rm GeV}^2$ to $9.4\,{\rm GeV}^2$ as a discrete sum, using the measured values of $\sigma(t)$ plus the 1$\sigma$  error. We find $|\delta a^p_k|\leq 0.013$.  The contribution above $9.4\,{\rm GeV}^2$ can be conservatively estimated by assuming a constant value for the form factors. This gives a constant value of $0.031$ for $\sqrt{ |G^p_M(t)|^2 + 2m_p^2|G^p_E(t)|^2/t}$ above $9.4\,{\rm GeV}^2$, leading to an additional $0.004$. In total we have $|\delta a^p_k|\leq 0.013+0.004$. 

Using the $e^+e^-\to n\bar n$ data from \cite{Antonelli:1998fv} for 
$t=3.61$ to $5.95\,{\rm GeV}^2$, we find $|\delta a^n_k|\leq 0.011$.  Assuming a constant value, $0.32$, for $\sqrt{ |G^n_M(t)|^2 + 2m_n^2|G^n_E(t)|^2/t}$ above $5.95\,{\rm GeV}^2$, leads to an additional $0.047$, giving in total $|\delta a^n_k|\leq 0.011+0.047$. 

These results are very similar to those of the electric form factor in \cite{Hill:2010yb}, although for the electric form factor more stringent bounds were obtained. Compared to the bounds calculated above, these contributions are negligible.  Our conclusion, as in \cite{Hill:2010yb}, is that the contribution of the physical timelike region $t\geq 4m_N^2$ can be neglected.

\subsubsection{Summary}
All our studies point out that for the magnetic form factor the coefficients  $a_k$ are smaller than 10. Since $a_0=G^{(1)}(0)=\mu_p-\mu_n\approx 4.7$, a bound of 5 might be too stringent.  In the following we will use a bounds of 10 and 15 instead of the bounds of 5 and 10 used in \cite{Hill:2010yb}. As we will see, even using a bound of 20 will not change the results in an appreciable way.  

One could also argue that a bound on the \emph{ratio}  $|a_k/a_0|\leq 5,10$ is more appropriate. Since $a_0$ is known, this will translate to a bound of  $|a_k|\leq 25,50$ in the $I=1$ case. We prefer to use the more stringent bound of $|a_k|\leq 10,15$, but we will comment on the results when using  these looser bounds.

It should be noted that for $t_0=0$, the magnetic radius depends only the coefficient of $z$. Writing  $G_{M}^p(q^2) = \sum_{k=0}^\infty a_k \, z(q^2)^k$, where $z(q^2)\equiv z(q^2,4m_\pi^2,0)$, equation (\ref{definition}) implies that 
\begin{equation}
r_M^p=\dfrac{\hbar c}{2m_\pi c^2}\sqrt{-\dfrac{3^{}a_1}{2\mu_p}},
\end{equation}
 where we are showing explicitly the factors of $\hbar$ and $c$.  A bound of $5,\,10,\,15$, or 20, on $|a_k|$,  implies also a  bound of 1.2, 1.6, 2.0, and 2.3 fm on $r_M^p$. Writing $G_M^{(0)}(q^2)= \sum_{k=0}^\infty a^{(0)}_k \, z(q^2,9m_\pi^2,0)^k$ and $G_M^{(1)} (q^2)=\sum_{k=0}^\infty a^{(1)}_k \, z(q^2,4m_\pi^2,0)^k$ we have 
 \begin{equation}\label{rmpiso}
 r_M^p=\dfrac{\hbar c}{2m_\pi c^2}\sqrt{-\dfrac{a^{(0)}_1+\frac94a^{(1)}_1}{3\mu_p}}. 
 \end{equation}
 A bound of 5, 10, 15, or 20, on $|a^{(0,1)}_k|$,  implies also a  bound of 0.98, 1.4, 1.7, or 2.0 fm on $r_M^p$. For our default choice of bounds of 10 and 15 these values are much larger than the current range of values quoted by the PDG  \cite{Beringer:1900zz}, roughly $0.7-0.9$ fm . Thus just the presence of our default bounds does not bias the extraction of the radius.

\section{Extraction of the proton magnetic radius}\label{proton}
\subsection{Proton data}
We extract the proton magnetic radius from the values of $G_M^p$ tabulated in \cite{Arrington:2007ux}. We write the  form factor as $G_M^p(q^2) = \sum_{k=0}^\infty a_k \, z(q^2)^k$, where  $z(q^2)\equiv z(q^2,4m_\pi^2,0)$. We fit $k<k_{\rm max}$ parameters, where $k_{\rm max}=2,\dots,12$. We minimize the $\chi^2$ function 
\begin{equation}\label{chi2}
\chi^2=\sum_i(\mbox{data}_{\,i}- \mbox{theory}_i)^2/(\sigma_i)^2,
\end{equation}
Where $i$ ranges over the tabulated values of \cite{Arrington:2007ux} up to a given maximal value of $Q^2$, with $Q^2=0.1,0.2,\dots,1.2,1.4,1.6,1.8$ GeV$^2$.  As explained above, our default choice for the bounds on the coefficients is $|a_k|<10$ and $|a_k|<15$.
 The proton magnetic radius is obtained from (\ref{definition}). The error bars are determined from the $\Delta\chi^2=1$ range.  Usually, the $\Delta\chi^2=1$ range was determined from a numerical search algorithm. For some higher values of $Q^2$,  the $\chi^2(r_M^p)$ seems to have some discontinuities and in that case, the $\Delta\chi^2=1$ was extracted directly from the $\chi^2(r_M^p)$ curve. To ensure a conservative estimate of the error, we quote only one digit in the error bar. 
 
The extracted values and the value of the minimum of $\chi^2$ do not vary with $k_{\rm max}$ for $k_{\rm max}>4$. In other words, the extracted values do not depend on the number of coefficients we fit.  In the following we quote results with $k_{\rm max}=8$.
 The extracted values of the magnetic radius are very consistent over the range of $Q^2$. Thus for data with $Q^2\leq 0.5$ GeV$^2$, we have  $r_M^p=0.91^{+0.03}_{-0.06}$ fm  for a bound of 10 and $r_M^p=0.92^{+0.04}_{-0.07}$ fm for a bound of 15, while for  $Q^2\leq 1.0$ GeV$^2$ we have $r_M^p=0.90^{+0.03}_{-0.07}$ fm  for a bound of 10 and $r_M^p=0.91^{+0.04}_{-0.07}$ fm for a bound of 15. 
 
We have studied the dependance of the extracted magnetic radius on the bounds on $|a_k|$. If we use a bound of $|a_k|<20$, the results above change to  $r_M^p=0.93^{+0.03}_{-0.07}$  fm for $Q^2\leq 0.5$ GeV$^2$ and  $r_M^p=0.91^{+0.04}_{-0.08}$  fm for $Q^2\leq 1.0$ GeV$^2$. These values are very similar to the ones obtained with $|a_k|<10$ and $|a_k|<15$.  As discussed above we consider the bound  $|a_k|<5$ to be too stringent, but if we do use it we obtain $r_M^p=0.89^{+0.03}_{-0.05}$  fm for $Q^2\leq 0.5$ GeV$^2$ and  $r_M^p=0.89^{+0.02}_{-0.05}$  fm for $Q^2\leq 1.0$ GeV$^2$, which are not statistically different from the results of our default bounds. 

Another possible choice of bounds might be to bound $|a_k/a_0|$. This is motivated by the fact that the vector dominance ansatz and the $\pi$-$\pi$ data indicate that $a_k/a_0$ is similar for the electric and magnetic form factors. Thus we might choose $|a_k/a_0|<5,10$. We have checked the effect of these looser bounds on the extracted magnetic radius.  For $Q^2\leq 0.5$ GeV$^2$, we have  $r_M^p=0.92^{+0.03}_{-0.07}$ fm  for a bound of $|a_k/a_0|<5$ and $r_M^p=0.95^{+0.04}_{-0.08}$ fm for a bound of $|a_k/a_0|<10$ while for  $Q^2\leq 1.0$ GeV$^2$ we have $r_M^p=0.91^{+0.04}_{-0.08}$ fm  for a bound of $|a_k/a_0|<5$ and $r_M^p=0.92^{+0.05}_{-0.09}$ fm for a bound of $|a_k/a_0|<10$. For the magnetic radius with $t_0=0$, $a_0=\mu_p\approx 2.8$, so if we choose  $|a_k/a_0|<5,10$ this translates to $|a_k|<14,28$ respectively. Comparing these results to the ones obtained above we notice a slight monotonic increase in the central value and the error bars with the loosening of the bound. The increase in the error bars is to be expected of course. Even with the looser bounds, the results we obtain are consistent with our default bounds. 

Using our default bounds of  $|a_k|<10$ and $|a_k|<15$, and using $Q^2\leq 0.5$ GeV$^2$ for concreteness we obtain $r_M^p=0.91^{+0.03}_{-0.06}\pm0.02 $ fm. The first error is for a bound of 10 and the second error includes the maximum variation of the $\Delta\chi^2=1$ interval when we redo the fits with a bound of 15.

 \subsection{Proton and neutron data}
Including neutron data allows us to separate the $I=1$ and $I=0$  isospin components of the proton magnetic form factor. Since for the $I=0$ components $t_{\rm cut}=9m_\pi^2$, this increases the value of $t_{\rm cut}$ and effectively decreases the maximum value of $z$. 

As before we use values of $G_M^p$ tabulated in \cite{Arrington:2007ux}. For $G_M^n(Q^2)$ we use values published in \cite{Lung:1992bu,Gao:1994ud, Anklin:1994ae,Anklin:1998ae, Kubon:2001rj, Anderson:2006jp, Lachniet:2008qf}\footnote{\cite{Anderson:2006jp} contain the final results that supersedes the previous publications \cite{Xu:2000xw, Xu:2002xc}. For \cite{Lachniet:2008qf}, the data is tabulated in \cite{CLAS}.}. We do not use the data reported in \cite {Markowitz:1993hx} and \cite{Bruins:1995ns}, as they were criticized for missing a systematic error, see section VIII of  \cite{Anderson:2006jp}\footnote{If we include these additional data points we obtain similar values of the magnetic radius but with much larger values of $\chi^2$.}.

We form the $\chi^2$ as before and express $G_M^n$ and $G_M^p$ in terms of $G_M^{(0)}$ and $G_M^{(1)}$, see (\ref{iso}). We express $G_M^{(0)}$  as a power series in $z(t,9m_\pi^2,0)$ and $G_M^{(1)}$ as a power series in $z(t,4m_\pi^2,0)$, i.e.
\begin{eqnarray}
G_M^{(0)}(t)&=&\sum_k a^{(0)}_kz^k(t, t_{\rm cut}=9m_\pi^2,0)\\
G_M^{(1)}(t)&=&\sum_k a^{(1)}_kz^k(t, t_{\rm cut}=4m_\pi^2,0)\,. 
\end{eqnarray}  

 As for the proton data alone, the extracted values of the magnetic radius do not depend on the number of the parameters we fit. The values are very consistent over the range of $Q^2$. Thus for data with $Q^2\leq 0.5$ GeV$^2$, we have  $r_M^p=0.87^{+0.04}_{-0.05}$ fm  for a bound of 10 and $r_M^p=0.87^{+0.05}_{-0.05}$ fm for a bound of 15, while for  $Q^2\leq 1.0$ GeV$^2$ we have $r_M^p=0.87^{+0.03}_{-0.05}$ fm  for a bound of 10 and $r_M^p=0.88^{+0.04}_{-0.05}$ fm for a bound of 15. These values are consistent with the values extracted from the proton data alone.  
 
We have studied the dependance of the extracted magnetic radius on the bounds on $|a_k|$. If we use a bound of $|a_k|<20$, the results above change to  $r_M^p=0.88^{+0.04}_{-0.06}$  fm for $Q^2\leq 0.5$ GeV$^2$ and  $r_M^p=0.88^{+0.05}_{-0.06}$  fm for $Q^2\leq 1.0$ GeV$^2$. These values are very similar to the ones obtained with $|a_k|<10$ and $|a_k|<15$.  If we use the bound $|a_k|<5$, we obtain $r_M^p=0.87^{+0.02}_{-0.02}$  fm for $Q^2\leq 0.5$ GeV$^2$ and  $r_M^p=0.87^{+0.02}_{-0.02}$  fm for $Q^2\leq 1.0$ GeV$^2$. The central values are consistent with our default bounds, but the error bars are substantially smaller. This is to be expected since this bound is too stringent. 

As explained above, another possible choice of bounds is $|a_k/a_0|<5,10$.  For $Q^2\leq 0.5$ GeV$^2$, we have in this case $r_M^p=0.88^{+0.05}_{-0.06}$ fm for a bound of $|a_k/a_0|<5$ and $r_M^p=0.91^{+0.05}_{-0.07}$ fm for a bound of $|a_k/a_0|<10$. For  $Q^2\leq 1.0$ GeV$^2$ we have $r_M^p=0.89^{+0.04}_{-0.07}$ fm  for a bound of $|a_k/a_0|<5$ and $r_M^p=0.90^{+0.05}_{-0.09}$ fm for a bound of $|a_k/a_0|<10$. Since $a^{(0)}_0=\mu_p+\mu_n\approx 0.88$, $a^{(1)}_0=\mu_p-\mu_n\approx 4.7$, $|a^{(0)}_k/a^{(0)}_0|<5$ implies $|a^{(0)}_k|<4.4$ and $|a^{(1)}_k/a^{(1)}_0|<5$ implies  $|a^{(1)}_k|<23.5$. Similarly $|a^{(0)}_k/a^{(0)}_0|<10$ implies  $|a^{(0)}_k|<8.8$  and $|a^{(1)}_k/a^{(1)}_0|<10$ implies  $|a^{(1)}_k|<47$. Comparing these results to the ones obtained above we notice again a monotonic increase in the central value and the error bars with the loosening of the bound. The increase in the error bars is to be expected of course. Even with the looser bounds, the results we obtain are consistent.  

Using our default bounds of  $|a_k|<10$ and $|a_k|<15$, and using $Q^2\leq 0.5$ GeV$^2$ for concreteness we obtain $r_M^p=0.87^{+0.04}_{-0.05}\pm0.01 $ fm.

\subsection{Proton, neutron, and $\pi\pi$ data} 
Between the two-pion and four-pion threshold the only state that can contribute to the imaginary part of the magnetic isovector form factor is that of two pions. Since we have information about ${\rm Im}\, G_M^{(1)}(t)$ in this region, see (\ref{pipiNN}), we can use it to raise the effective threshold for the isovector form factor from $t_{\rm cut}=4m_\pi^2$ to  $t_{\rm cut}=16m_\pi^2$.  We do that by fitting \cite{Hill:2010yb} 
\begin{equation}\label{Gcut}
G_M^{(1)}(t)=G_{\rm cut}(t)+\sum_k a^{(1)}_kz^k(t, t_{\rm cut}=16m_\pi^2,0). 
\end{equation}
$G_{\rm cut}(t)$ is calculated using (\ref{dispersion}) from the discrete expression for  ${\rm Im}\, G_M^{(1)}(t)$ described in section \ref{sec232}. As in  \cite{Hill:2010yb} 
we consider two cases for $G_{\rm cut}(t)$. The first is generated by the values of ${\rm Im}\, G_M^{(1)}(t)$ in the range $4m_\pi^2<t<16m_\pi^2$, and the second by the values of ${\rm Im}\, G_M^{(1)}(t)$ in the range $4m_\pi^2<t<40m_\pi^2$. The second choice amounts to modeling the $\pi\pi$ continuum $16m_\pi^2<t<40m_\pi^2$ by ${\rm Im}\, G_M^{(1)}(t)$ of (\ref{pipiNN}). As explained in \cite{Hill:2010yb}, this does not introduce model dependance since the difference between the true continuum and $G_{\rm cut}(t)$ will be accounted for by the parameters in the $z$ expansion, as we do not change the value of $t_{\rm cut}=16m_\pi^2$. 

In \cite{Hill:2010yb} it was found that the second choice of $G_{\rm cut}(t)$ led to a smaller size of the coefficients in the $z$ expansion of the isovector form factor. We would like to check if that holds true also in  the magnetic case. We fit the same proton and neutron data for $Q^2_{\rm max}=1$ GeV$^2$, $t_0=0$, $k_{\rm max}=8$ and a bound of 15 on the coefficients using (\ref{Gcut}). For the first choice of $G_{\rm cut}(t)$ we find the first two coefficients of the isoscalar form factor to be  $-2^{+0.5}_{-0.3}, 3^{+2}_{-6}$ and the first two coefficients of the vector form factor to be  $-13.5(3), 13^{+6}_{-3}$ (the value of $13^{+6}_{-3}$ was obtained by applying a bound of 15 on all the coefficients  with the exception of the second one, which is left unbounded). For the second choice of $G_{\rm cut}(t)$ we find the first two coefficients of the isoscalar form factor are not changed while the first two coefficients of the vector form factor are  $2.6^{+0.4}_{-0.5}, 5^{+5}_{-4}$. As in the electric form factor case, we have a reduction in the size of the isovector coefficients when using the second form. We will therefore adopt that as our default choice. As we will see below, the value of the magnetic radius does not change if we use the first form of  $G_{\rm cut}(t)$. 

We can understand the large size of the isovector coefficients when using  $G_{\rm cut}(t)$ calculated from ${\rm Im}\, G_M^{(1)}(t)$ in the range $4m_\pi^2<t<16m_\pi^2$. From equations (\ref{iso}) and (\ref{Gcut}), the proton magnetic radius is given by 
 \begin{equation}
 r_M^p=\dfrac{\hbar c}{2m_\pi c^2}\sqrt{\dfrac1{\mu_p}\left(-\dfrac{1}{3}a^{(0)}_1-\frac{3}{16}a^{(1)}_1+4m_\pi^2c^4G^\prime_{\rm cut}(0)\right)},
 \end{equation}
where $G^\prime_{\rm cut}(0)$ is obtained from (\ref{dispersion})
\begin{equation}\label{Gcutprime}
G^\prime_{\rm cut}(0) = {1\over \pi} \int_{4m_\pi^2}{dt^\prime}\, {{\rm Im}G(t^\prime + i0) \over (t^\prime)^2 } \,. 
\end{equation}
Since ${\rm Im}\, G_M^{(1)}(t)$ from (\ref{pipiNN}) is positive in the relevant region, as we increase the upper limit in (\ref{Gcutprime}), $G^\prime_{\rm cut}(0)$ increases.  Therefore $G^\prime_{\rm cut}(0)$ calculated from ${\rm Im}\, G_M^{(1)}(t)$ in the range $4m_\pi^2<t<16m_\pi^2$ is smaller than $G^\prime_{\rm cut}(0)$ calculated from ${\rm Im}\, G_M^{(1)}(t)$ in the range $4m_\pi^2<t<40m_\pi^2$ and as a result $|a^{(1)}_1|$ must be larger to maintain the same size of $r_M^p$ preferred by the data. In fact, since we can calculate $G^\prime_{\rm cut}(0)$, if we \emph{assume} $r_M^p\approx 0.87$  fm and use  $a^{(0)}_1 \approx -2$,  we can calculate and find $a^{(1)}_1\approx -13$ in the first case and $a^{(1)}_1\approx 3$ in the second case. These are the values we obtained above 

Using (\ref{Gcut}) we extract the magnetic radius.  The extracted values of the magnetic radius do not depend on the number of the parameters we fit. The values are very consistent over the range of $Q^2$. Thus for data with $Q^2\leq 0.5$ GeV$^2$, we have  $r_M^p=0.871^{+0.011}_{-0.015}$ fm  for a bound of 10 and $r_M^p=0.873^{+0.012}_{-0.016}$ fm for a bound of 15, while for  $Q^2\leq 1.0$ GeV$^2$ we have $r_M^p=0.874^{+0.008}_{-0.015}$ fm  for a bound of 10 and $r_M^p=0.874^{+0.012}_{-0.014}$ fm for a bound of 15. These values are consistent with the values extracted above. 

We have studied the dependance of the radius on the bounds on the coefficients. If we use a bound of 20, we have $r_M^p=0.876^{+0.012}_{-0.018}$ for $Q^2\leq 0.5$ GeV$^2$ and  $r_M^p=0.875^{+0.013}_{-0.016}$ for $Q^2\leq 1.0$ GeV$^2$. These values are very similar to the ones we obtain with a bound of 10 and 15. If we use the too-stringent bound of 5 we obtain $r_M^p=0.867^{+0.010}_{-0.013}$ for $Q^2\leq 0.5$ GeV$^2$ and  $r_M^p=0.867^{+0.006}_{-0.008}$ for $Q^2\leq 1.0$ GeV$^2$. These values are consistent, but the error bars are smaller. 

Another possible choice of bounds is $|a_k/a_0|<5,10$. For $Q^2\leq 0.5$ GeV$^2$, we find $r_M^p=0.867^{+0.013}_{-0.013}$ fm for a bound of $|a_k/a_0|<5$ and $r_M^p=0.869^{+0.013}_{-0.015}$ fm for a bound of $|a_k/a_0|<10$. For $Q^2\leq 1.0$ GeV$^2$, we find $r_M^p=0.867^{+0.008}_{-0.009}$ fm for a bound of $|a_k/a_0|<5$ and $r_M^p=0.873^{+0.009}_{-0.014}$ fm for a bound of $|a_k/a_0|<10$. All these results are consistent with our default  choices. 

The decrease in the error bars when including the $\pi\pi$ data arises from the increase in the value of $t_{\rm cut}$ from $4m_\pi^2$ to $16m_\pi^2$ for the isovector form factor.  If we use (\ref{Gcut}) but with $t_{\rm cut}=4m_\pi^2$ we obtain results that are almost identical to the fits using the proton and neutron data alone.  As another check of our results, we fit the data using (\ref{Gcut}), but with $G_{\rm cut}(t)$ calculated using ${\rm Im}\, G_M^{(1)}(t)$ in the range $4m_\pi^2<t<16m_\pi^2$. As discussed above, we use only a bound of 15 in this case. For $Q^2\leq 0.5$ GeV$^2$  we find $r_M^p=0.873^{+0.011}_{-0.016}$, and for $Q^2\leq 1.0$ GeV$^2$  we find $r_M^p=0.873^{+0.012}_{-0.012}$. These values are very close to the ones we obtained with the use of the default form of $G_{\rm cut}(t)$. 

The expression for ${\rm Im}\, G_M^{(1)}(t)$ depends on $f^1_-(t)$. The tabulation of $f^1_-(t)$ in \cite{Hohler} does not quote any error.  In  \cite{Hill:2010yb} an error of 30\% was used as a representative uncertainty. If we assume a 30\%  increase  for $f^1_-(t)$ and hence for $G_{\rm cut}(t)$ we obtain for  $Q^2\leq 0.5$ GeV$^2$ and a bound of 10, $r_M^p=0.872^{+0.013}_{-0.015}$. If we assume a 30\% decrease for $G_{\rm cut}(t)$  we obtain for $Q^2\leq 0.5$ GeV$^2$ and a bound of 10, $r_M^p=0.867^{+0.010}_{-0.015}$.

In summary, all our checks produce consistent results for $r_M^p$. Using our default choices for the bounds and $G_{\rm cut}(t)$, and using $Q^2\leq 0.5$ GeV$^2$ for concreteness we obtain $r_M^p=0.87^{+0.02}_{-0.02} $ fm. Our conservative error estimate includes  the variation of the bounds and of $G_{\rm cut}(t)$ where we choose to quote only one digit in our error estimate. 

\section{Extraction of the neutron magnetic radius}\label{neutron}
The data we have used to extract the magnetic radius of the proton can be used also to extract the magnetic radius of the neutron. The magnetic radius of the neutron is defined as $r_M^n \equiv \sqrt{\langle r^2\rangle_M^n}$, where  

\begin{equation}
\langle r^2 \rangle_M^n=
\frac{6}{G_M^n(0)}\frac{d}{dq^2}G_M^n(q^2)\bigg|_{q^2=0}\,.
\end{equation}
We extract the neutron magnetic radius from the neutron, neutron and proton, and neutron, proton, and $\pi\pi$ data sets. We follow the same default choices described above. In particular we will use a bound of 10 and 15 on the coefficients of the $z$ expansion. \\

\subsection{Neutron data}
 
Using the neutron form factor data reported in  \cite{Lung:1992bu,Gao:1994ud, Anklin:1994ae,Anklin:1998ae, Kubon:2001rj, Anderson:2006jp, Lachniet:2008qf} we fit $G_M^n(q^2) = \sum_{k=0}^\infty a_k \, z(q^2)^k$ by minimizing the $\chi^2$ function of (\ref{chi2}). For a cut $Q^2\leq 0.5$ GeV$^2$ we find $r_M^n=0.74^{+0.13}_{-0.06}$ fm for a bound of 10 and $r_M^n=0.65^{+0.21}_{-0.07}$ fm for a bound of 15. For a cut of $Q^2\leq 1.0$ GeV$^2$ we find $r_M^n=0.77^{+0.17}_{-0.09}$ fm for a bound of 10 and $r_M^n=0.74^{+0.20}_{-0.11}$ fm for a bound of 15. Obviously the error bars for $r_M^n$ extracted from the neutron data are much larger than for $r_M^p$. We prefer to quote only one digit in our error bar. We  therefore determine  $r_M^n=0.7^{+0.2}_{-0.1}$ fm from neutron data alone.  Comparing to $r_M^p=0.91^{+0.03}_{-0.06}\pm0.02$ fm obtained from proton data alone, we find that $r_M^n$ and $r_M^p$ are consistent within errors. 
 \subsection{Neutron and proton data}

Adding the proton form factor data from  \cite{Arrington:2007ux} allows us to separate the isospin components. The magnetic radius of the neutron is given by an equation similar to  (\ref{rmpiso})
\begin{equation}
 r_M^n=\dfrac{\hbar c}{2m_\pi c^2}\sqrt{\dfrac{-a^{(0)}_1+\frac94a^{(1)}_1}{3\mu_n}}. 
 \end{equation}
We fit  the isoscalar and the isovector form factors as described before.  For a cut $Q^2\leq 0.5$ GeV$^2$ we find $r_M^n=0.89^{+0.06}_{-0.09}$ fm for a bound of 10 and $r_M^n=0.88^{+0.08}_{-0.09}$ fm for a bound of 15. For a cut of $Q^2\leq 1.0$ GeV$^2$ we find $r_M^n=0.88^{+0.06}_{-0.08}$ fm for a bound of 10 and $r_M^n=0.89^{+0.07}_{-0.10}$ fm for a bound of 15. Again the error bars for $r_M^n$ are about twice as large as those for $r_M^p$ from the same data set. Quoting only one digit  we  determine  $r_M^n=0.9^{+0.1}_{-0.1}$ fm from neutron and proton data. Comparing to $r_M^p=0.87^{+0.04}_{-0.05}\pm0.02$ fm obtained from the same proton and neutron data, we find that $r_M^n$ and $r_M^p$ are consistent within errors.

\subsection{Neutron, proton, and $\pi\pi$ data} 
Adding the $\pi\pi$ data as described in the previous section leads to a reduction in the error bars. For a cut $Q^2\leq 0.5$ GeV$^2$ we find $r_M^n=0.89^{+0.03}_{-0.03}$ fm for a bound of 10 and $r_M^n=0.89^{+0.03}_{-0.03}$ fm for a bound of 15. If we take a 30\% variation of $f^1_-(t)$ as described above, we get values of  $r_M^n$ within this range. For a cut of $Q^2\leq 1.0$ GeV$^2$ we find $r_M^n=0.88^{+0.03}_{-0.01}$ fm for a bound of 10 and $r_M^n=0.88^{+0.03}_{-0.02}$ fm for a bound of 15. As before the error bars for $r_M^n$ are about twice as large as those for $r_M^p$ from the same data set. Quoting only one digit  for the error bars we  determine  $r_M^n=0.89^{+0.03}_{-0.03}$ fm from neutron, proton, and $\pi\pi$ data. Comparing to $r_M^p=0.87^{+0.02}_{-0.02}$ fm obtained from the same data set, we find that $r_M^n$ and $r_M^p$ are consistent within errors.

\section{Conclusions}\label{conclusions}
The recent large discrepancy in the extraction of the charge radius of the proton from spectroscopic measurements  of regular and muonic hydrogen has motivated the reexamination of the extraction of nucleon radii from scattering data. Since the first measurement of the ``size" of the proton \cite{Mcallister:1956ng} almost 60 years ago, there have been many extractions of the charge radius of the proton. These were based on different data sets and postulated different functional forms for the form factors.   These various extractions do not agree with each other. Even when using the same data sets, different functional forms can lead to different values of the charge radius of the proton.  

A fundamental problem of many of these extractions is that they do not take into account the known analytic structure of the form factors. Therefore, it is unlikely that an arbitrary functional form will be consistent with this structure. This analytic structure constrains the form factors but does not determine it completely.  Since the form factors are non-perturbative functions, one would like to incorporate the analytic structure while maintaining the flexibility of the functional form. The so-called ``$z$ expansion" described in the introduction achieves both of these goals. It automatically incorporates the analytic structure and allows  for flexible functional forms. It is therefore not surprising that the $z$ expansion has become a standard tool in analyzing meson form factors; see for example section 8.3.1 of \cite{Aoki:2013ldr}. 

To the best of our knowledge the first application of the $z$ expansion to baryon form factor was done in \cite{Hill:2010yb}. That paper also has shown the need to impose some constraints on the coefficients of the $z$ expansion in order to have a result that is independent of the number of parameters. For meson form factors such as $B\to \pi$, constraints that bound the sum of the squares of the coefficients can be obtained from unitarity\footnote{See \cite{Becher:2005bg} for a discussion of these unitarity bounds.}. For the nucleon form factors such constraints are less useful since there is a large distance between the two-pion threshold where the singularity begins, and the two-nucleon threshold where the unitarity bounds can be applied.  The studies of \cite{Hill:2010yb} have shown that a uniform bound on the coefficients can be applied. The methods of \cite{Hill:2010yb} were later used in \cite{Bhattacharya:2011ah} for a model-independent extraction of  the axial mass parameter of the nucleon from neutrino-nucleon scattering data. 

We have applied the same methods in this paper to extract the magnetic radius of the proton from scattering data in a model-independent way. While not as severe as the proton charge radius problem, various extractions in recent years, e.g the ones cited by the PDG \cite{Beringer:1900zz}, are not consistent with each other.  The goal of our study was to try and resolve these discrepancies.  

We first studied the bounds on the coefficients of the $z$ expansion. In  \cite{Hill:2010yb} bounds of 5 and 10 were used. Since the value of the isovector magnetic form factor at zero momentum transfer is about 4.7, a bound of 5 on the coefficients might be too stringent. Our studies have shown that this is indeed the case, but bounds of 10 and 15 are conservative enough for the coefficients of the magnetic form factor. An alternative option is to use a bound of 5  and 10 on the ratio  $|a_k/a_0|$. Fitting the data using each of these prescriptions gives consistent results. Our default choice is to use the bound of 10 and 15. 

We have extracted the magnetic radius of the proton from  three data sets. The first contains values of proton magnetic form factor data  tabulated in  \cite{Arrington:2007ux}. The second contains the proton data and the neutron magnetic form factor data  tabulated in \cite{Lung:1992bu,Gao:1994ud, Anklin:1994ae,Anklin:1998ae, Kubon:2001rj, Anderson:2006jp, Lachniet:2008qf}. The third contains the proton and neutron data and the two-pion continuum data constructed from pion form factor data and a $\pi\pi\to N\bar N$ partial amplitude  using  (\ref{pipiNN}). In all the cases we use the listed data and do not apply any corrections. For each data set the extracted magnetic radius of the proton is  consistent as we change  the number of parameters we fit, or the cut on $Q^2$. Taking $Q^2\leq0.5$ GeV$^2$ and fits with eight parameters for concreteness, we find that for the proton data set $r_M^p=0.91^{+0.03}_{-0.06}\pm0.02 $ fm, for the proton and neutron data set $r_M^p=0.87^{+0.04}_{-0.05}\pm0.01$ fm, and for the proton, neutron, and $\pi\pi$ data set  $r_M^p=0.87^{+0.02}_{-0.02} $ fm. For the first two values the first error is for a bound of 10 and the second error includes the maximum variation of the $\Delta\chi^2=1$ interval when we redo the fits with a bound of 15. The error on the third value combines both, as well as errors on the continuum contribution as discussed in section \ref{proton}. In all cases we choose to quote one digit in our error bar.  As expected the errors decrease as we include more data, but the main effect is the change in the value of  $t_{\rm cut}$.  Using proton data alone we have $t_{\rm cut}=4m_\pi^2$. Adding the neutron data allows to set  $t_{\rm cut}=9m_\pi^2$ for the isoscalar magnetic form factor. Adding the two-pion continuum allows to set $t_{\rm cut}=16m_\pi^2$ for the isovector magnetic form factor. The increase in $t_{\rm cut}$ leads to a decrease in the maximum value of $|z|$ and therefore for a smaller error. 

Comparing our third value of the magnetic radius of the proton, $r_M^p=0.87^{+0.02}_{-0.02} $ fm, to the values quoted by the PDG \cite{Beringer:1900zz}, we find that they are more consistent with $r_M^p=0.876\pm0.010\pm0.016$ fm extracted in \cite{Borisyuk:2009mg}  and  $r_M^p=0.854 \pm 0.005$ fm extracted in \cite{Belushkin:2006qa}, than the A1 collaboration value of  $r_M^p=0.777(13)_{\rm stat.}(9)_{\rm syst.}(5)_{\rm model}(2)_{\rm group}$ fm \cite{Bernauer:2010wm}. Our error bars are larger than these extraction, which is not unusual when using model-independent methods \cite{Hill:2010yb, Bhattacharya:2011ah}.   Other extractions of the proton magnetic radius from scattering data that were not quoted by the PDG are $r^p_M=0.855\pm 0.035$ fm \cite{Sick:2005az}, $r^p_M=0.867\pm 0.020$ fm \cite{Zhan:2011ji}, and $r^p_M=0.86^{+0.02}_{-0.03}$ fm \cite{Lorenz:2012tm}. Our results are consistent with theses values too. 

The same data can be used also for a model-independent extraction of the neutron magnetic radius.  Taking $Q^2\leq0.5$ GeV$^2$ and fits with eight parameters for concreteness we find that for the neutron data set $r_M^n=0.7^{+0.2}_{-0.1}$ fm, for the proton and neutron data set $r_M^n=0.9^{+0.1}_{-0.1} $ fm, and for the proton, neutron, and $\pi\pi$ data set  $r_M^n=0.89^{+0.03}_{-0.03} $ fm. The last value can be compared to the value quoted by the PDG, $r_M^n=0.862^{+0.009}_{-0.008} $ fm \cite{Belushkin:2006qa}. Our results are consistent but our error bars are  larger which can again be attributed to the use of model-independent methods.  

It is interesting to note that the magnetic radius of the neutron is consistent within errors with the magnetic radius of the proton. In fact the magnetic radius of the proton is also consistent within errors with the value of the charge radius of the proton, $r_E^p=0.871\pm0.009\pm0.002\pm0.002$ fm extracted using the same model-independent methods in \cite{Hill:2010yb}. The use of model-independent methods is essential in establishing these facts. Similar results were obtained in \cite{Belushkin:2006qa}. That paper also utilized the analytic properties of the form factors, but it postulates a specific functional form for them, unlike our more general  $z$ expansion. 

It is well known that the ratio of the proton and neutron magnetic moments can be explained by using SU(6) symmetry  \cite{Beg:1964nm}  or the quark model \cite{Becchi:1965zza}. It is therefore not inconceivable that not just the magnetic moments (values of $G_M^p$ and $G_M^n$ at $q^2=0$) but the magnetic radii  (slopes of $G_M^p$ and $G_M^n$ at $q^2=0$) are related by some symmetry. We  leave that to a future study.  

Our study shows the utility and robustness of the $z$ expansion in model-independent extraction of fundamental properties of nucleons such as the electric, magnetic, and axial radii. It would be interesting to apply the same methods to newer data sets such as that of the A1 collaboration and to include also polarization data. These can lead to definitive and model-independent values for the proton magnetic and charge radii from scattering data.

%It is interesting to note that the magnetic radius of the neutron is consistent within errors with the magnetic radius of the proton. In fact the magnetic radius of the proton is also consistent within errors with the value of the charge radius of the proton, $r_E^p=0.871\pm0.009\pm0.002\pm0.002$ fm extracted using the same model-independent methods in    \cite{Hill:2010yb}. We will not interpret these results  here, but using model-independent methods is essential in establishing these facts. 

%Our study shows the utility and robustness of the $z$ expansion in model-independent extraction of fundamental properties of nucleons such as the electric, magnetic, and axial radii. It would be interesting to apply the same methods to newer data sets such as that of the A1 collaboration and to include also polarization data. 

\vskip 0.2in
\noindent
{\bf Acknowledgements}
\vskip 0.1in
\noindent
 We thank R.J. Hill and J.~Arrington for discussions and comments on the manuscript. This work was supported by NSF Grant PHY-1156651 (Z.E.), DOE grant DE-FG02-13ER41997 (G.P. and J.R.) and the NIST Precision Measurement Grants Program (G.P.)


\begin{thebibliography}{99}

\bibitem{Frisch}
R. Frisch and O. Stern, Zeitschrift f\"ur Physik {\bf 85}, 4 (1933).

\bibitem{Stern}
\url{http://www.nobelprize.org/nobel_prizes/physics/laureates/1943/stern-lecture.pdf}

\bibitem{Green:2014xba} 
  J.~R.~Green, J.~W.~Negele, A.~V.~Pochinsky, S.~N.~Syritsyn, M.~Engelhardt and S.~Krieg,
  %``Nucleon electromagnetic form factors from lattice QCD using a nearly physical pion mass,''
  arXiv:1404.4029 [hep-lat].


\bibitem{Pohl:2010zz}
  R.~Pohl {\it et al.},
  %``The size of the proton,''
  Nature {\bf 466}, 213 (2010).
  %%CITATION = NATUA,466,213;%%
  
\bibitem{Antognini:1900ns} 
  A.~Antognini, F.~Nez, K.~Schuhmann, F.~D.~Amaro, FrancoisBiraben, J.~M.~R.~Cardoso, D.~S.~Covita and A.~Dax {\it et al.},
  %``Proton Structure from the Measurement of $2S-2P$ Transition Frequencies of Muonic Hydrogen,''
  Science {\bf 339}, 417 (2013).
  %%CITATION = SCIEA,339,417;%%  


\bibitem{Mohr:2012tt} 
  P.~J.~Mohr, B.~N.~Taylor and D.~B.~Newell,
  %``CODATA Recommended Values of the Fundamental Physical Constants: 2010,''
  Rev.\ Mod.\ Phys.\  {\bf 84}, 1527 (2012)
  [arXiv:1203.5425 [physics.atom-ph]].
  %%CITATION = ARXIV:1203.5425;%%
  
 \bibitem{Daza:2010rh} 
  F.~Garcia Daza, N.~G.~Kelkar and M.~Nowakowski,
  %``Breit Equation with Form Factors in the Hydrogen Atom,''
  J.\ Phys.\ G {\bf 39}, 035103 (2012)
  [arXiv:1008.4384 [hep-ph]].
  
\bibitem{Jentschura:2010ej} 
  U.~D.~Jentschura,
  %``Lamb Shift in Muonic Hydrogen. I. Verification and Update of Theoretical Predictions,''
  Annals Phys.\  {\bf 326}, 500 (2011)
  [arXiv:1011.5275 [hep-ph]].  
  
  \bibitem{Hill:2011wy}
  R.~J.~Hill, G.~Paz,
  %``Model independent analysis of proton structure for hydrogen-like bound states,''
 Phys. Rev. Lett.  {\bf 107},  160402 (2011),  [arXiv:1103.4617 [hep-ph]] 
  
  
  
\bibitem{Carroll:2011rv} 
  J.~D.~Carroll, A.~W.~Thomas, J.~Rafelski and G.~A.~Miller,
  %``Non-Perturbative Relativistic Calculation of the Muonic Hydrogen Spectrum,''
  Phys.\ Rev.\ A {\bf 84}, 012506 (2011)
  [arXiv:1104.2971 [physics.atom-ph]].  
  
\bibitem{Eides:2012ue} 
  M.~I.~Eides,
  %``Weak Interaction Contributions in Light Muonic Atoms,''
  Phys.\ Rev.\ A {\bf 85}, 034503 (2012)
  [arXiv:1201.2979 [physics.atom-ph]].  
  
\bibitem{Kelkar:2012hf} 
  N.~G.~Kelkar, F.~G.~Daza and M.~Nowakowski,
  %``Determining the Size of the Proton,''
  Nucl.\ Phys.\ B {\bf 864}, 382 (2012)
  [arXiv:1203.0581 [hep-ph]].  
  
\bibitem{Borie:2012zz} 
  E.~Borie,
  %``Lamb shift in light muonic atoms: Revisited,''
  Annals Phys.\  {\bf 327}, 733 (2012).
  %%CITATION = APNYA,327,733;%%  
  
\bibitem{Antognini:2012ofa} 
  A.~Antognini, F.~Kottmann, F.~Biraben, P.~Indelicato, F.~Nez and R.~Pohl,
  %``Theory of the 2S-2P Lamb shift and 2S hyperfine splitting in muonic hydrogen,''
  Annals Phys.\  {\bf 331}, 127 (2013)
  [arXiv:1208.2637 [physics.atom-ph]].
  %%CITATION = ARXIV:1208.2637;%%
  
\bibitem{Indelicato:2012pfa} 
  P.~Indelicato,
  %``Non-perturbative evaluation of some QED contributions to the muonic hydrogen $n=2$ Lamb shift and hyperfine structure,''
  Phys.\ Rev.\ A {\bf 87}, no. 2, 022501 (2013)
  [arXiv:1210.5828 [physics.atom-ph]].  
  
\bibitem{Graczyk:2013pca} 
  K.~M.~Graczyk,
  %``Comparison of Neural Network and Hadronic Model Predictions of Two-Photon Exchange Effect,''
  Phys.\ Rev.\ C {\bf 88}, 065205 (2013)
  [arXiv:1306.5991 [hep-ph]].

\bibitem{Chen:2013udl} 
  D.~-Y.~Chen and Y.~-B.~Dong,
  %``Two-photon exchange in the $\ell$ +p ? $\ell$+p process with a massive lepton,''
  Phys.\ Rev.\ C {\bf 87}, no. 4, 045209 (2013).  
  
\bibitem{Giannini:2013bra} 
  M.~M.~Giannini and E.~Santopinto,
  %``On the proton radius problem,''
  arXiv:1311.0319 [hep-ph].
  %%CITATION = ARXIV:1311.0319;%%  

\bibitem{Korzinin:2013uia} 
  E.~Y.~.Korzinin, V.~G.~Ivanov and S.~G.~Karshenboim,
  %``$?^2(Z?)^4m$ contributions to the Lamb shift and the fine structure in light muonic atoms,''
  Phys.\ Rev.\ D {\bf 88}, no. 12, 125019 (2013)
  [arXiv:1311.5784 [physics.atom-ph]].
  %%CITATION = ARXIV:1311.5784;%%
  
 \bibitem{Indelicato:2014mra} 
  P.~Indelicato, P.~J.~Mohr and J.~Sapirstein,
  %``Coordinate-space approach to vacuum polarization,''
  arXiv:1402.0439 [quant-ph]. 
  
 \bibitem{Faustov:2014pwa} 
  R.~N.~Faustov, A.~P.~Martynenko, G.~A.~Martynenko and V.~V.~Sorokin,
  %``Radiative nonrecoil nuclear finite size corrections of order $\alpha(Z\alpha)^5$ to the hyperfine splitting of S-states in muonic hydrogen,''
  Phys.\ Lett.\ B {\bf 733}, 354 (2014)
  [arXiv:1402.5825 [hep-ph]].
  %%CITATION = ARXIV:1402.5825;%% 

\bibitem{Karshenboim:2014maa} 
  S.~G.~Karshenboim,
  %``A self-consistent value of the electric radius of the proton from the Lamb shift in muonic hydrogen,''
  arXiv:1405.6039 [hep-ph].
  
%%%%%%%%%%%%%%%%%%%%%%%%%%%%    

\bibitem{Friedmann:2009mx} 
  T.~Friedmann,
  %``No Radial Excitations in Low Energy QCD. I. Diquarks and Classification of Mesons,''
  Eur.\ Phys.\ J.\ C {\bf 73}, 2298 (2013)
  [arXiv:0910.2229 [hep-ph]].
  
\bibitem{Friedmann:2009mz} 
  T.~Friedmann,
  %``No Radial Excitations in Low Energy QCD. II. The Shrinking Radius of Hadrons,''
  Eur.\ Phys.\ J.\ C {\bf 73}, 2299 (2013)
  [arXiv:0910.2231 [hep-ph]].  
  
\bibitem{DeRujula:2010dp} 
  A.~De Rujula,
  %``QED is not endangered by the proton's size,''
  Phys.\ Lett.\ B {\bf 693}, 555 (2010)
  [arXiv:1008.3861 [hep-ph]].  
  
\bibitem{Cloet:2010qa} 
  I.~C.~Cloet and G.~A.~Miller,
  %``Third Zemach Moment of the Proton,''
  Phys.\ Rev.\ C {\bf 83}, 012201 (2011)
  [arXiv:1008.4345 [hep-ph]].  

\bibitem{Vanderhaeghen:2010nd} 
  M.~Vanderhaeghen and T.~Walcher,
  %``Long Range Structure of the Nucleon,''
  Nucl.\ Phys.\ News {\bf 21}, 14 (2011)
  [arXiv:1008.4225 [hep-ph]].
  

\bibitem{DeRujula:2010ub} 
  A.~De Rujula,
  %``Comment on `The third Zemach moment of the proton', by Cloet and Miller,''
  arXiv:1008.4546 [hep-ph].
  %%CITATION = ARXIV:1008.4546;%%  
  
\bibitem{Kholmetskii:2010sx} 
  A.~Kholmetskii, O.~Missevitch and T.~Yarman,
  %``Pure bound field theory and structure of atomic energy levels,''
  arXiv:1010.2845 [physics.atom-ph].    
  
  
  
\bibitem{DeRujula:2010zk} 
  A.~De Rujula,
  %``QED confronts the radius of the proton,''
  Phys.\ Lett.\ B {\bf 697}, 26 (2011)
  [arXiv:1010.3421 [hep-ph]].
  
 \bibitem{Distler:2010zq} 
  M.~O.~Distler, J.~C.~Bernauer and T.~Walcher,
  %``The RMS Charge Radius of the Proton and Zemach Moments,''
  Phys.\ Lett.\ B {\bf 696}, 343 (2011)
  [arXiv:1011.1861 [nucl-th]]. 
  
 \bibitem{Miller:2011yw} 
  G.~A.~Miller, A.~W.~Thomas, J.~D.~Carroll and J.~Rafelski,
  %``Natural Resolution of the Proton Size Puzzle,''
  Phys.\ Rev.\ A {\bf 84}, 020101 (2011)
  [arXiv:1101.4073 [physics.atom-ph]].
 
  
\bibitem{Carlson:2011zd} 
  C.~E.~Carlson and M.~Vanderhaeghen,
  %``Higher order proton structure corrections to the Lamb shift in muonic hydrogen,''
  Phys.\ Rev.\ A {\bf 84}, 020102 (2011)
  [arXiv:1101.5965 [hep-ph]].  
  
  
  
 \bibitem{Pineda:2011xp} 
  A.~Pineda,
  %``Brief Review of the Theory of the Muonic Hydrogen Lamb Shift and the Proton Radius,''
  arXiv:1108.1263 [hep-ph]. 
  
 \bibitem{Wu:2011jm} 
  B.~Y.~Wu and C.~W.~Kao,
  %``The Third Zemach Moment and the Size of the Proton,''
  arXiv:1108.2968 [hep-ph]. 
  
\bibitem{Carlson:2011dz} 
  C.~E.~Carlson and M.~Vanderhaeghen,
  %``Constraining off-shell effects using low-energy Compton scattering,''
  arXiv:1109.3779 [physics.atom-ph].
  
\bibitem{Kholmetsky:2012zz} 
  A.~L.~Kholmetsky, O.~V.~Missevitch and T.~Yarman,
  %``Hyperfine spin-spin interaction and Zeeman effect in the pure bound field theory,''
  Eur.\ Phys.\ J.\ Plus {\bf 127}, 44 (2012).  
  
\bibitem{Karr:2012mfa} 
  J.~-P.~Karr and L.~Hilico,
  %``Why three-body physics do not solve the proton radius puzzle,''
  Phys.\ Rev.\ Lett.\  {\bf 109}, 103401 (2012)
  [arXiv:1205.0633 [physics.atom-ph]].  
  
\bibitem{Birse:2012eb} 
  M.~C.~Birse and J.~A.~McGovern,
  %``Proton polarisability contribution to the Lamb shift in muonic hydrogen at fourth order in chiral perturbation theory,''
  Eur.\ Phys.\ J.\ A {\bf 48}, 120 (2012)
  [arXiv:1206.3030 [hep-ph]].  
  
\bibitem{Miller:2012ht} 
  G.~A.~Miller, A.~W.~Thomas and J.~D.~Carroll,
  %``Nuclear Quasi-Elastic Electron Scattering Limits Nucleon Off-Mass Shell Properties,''
  Phys.\ Rev.\ C {\bf 86}, 065201 (2012)
  [arXiv:1207.0549 [nucl-th]].  
  
\bibitem{Miller:2012ne} 
  G.~A.~Miller,
  %``Proton Polarizability Contribution: Muonic Hydrogen Lamb Shift and Elastic Scattering,''
  Phys.\ Lett.\ B {\bf 718}, 1078 (2013)
  [arXiv:1209.4667 [nucl-th]].
  
 \bibitem{Greenberg:2012vn} 
  O.~W.~Greenberg and S.~Cowen,
  %``N-quantum calculation of the hydrogen atom with one-photon exchange,''
  Phys.\ Rev.\ A {\bf 87}, 042516 (2013)
  [arXiv:1211.1619 [quant-ph]]. 
  
\bibitem{Gorchtein:2013yga} 
  M.~Gorchtein, F.~J.~Llanes-Estrada and A.~P.~Szczepaniak,
  %``$\mu-H$ Lamb shift: dispersing the nucleon-excitation uncertainty with a finite energy sum rule,''
  Phys.\ Rev.\ A {\bf 87}, 052501 (2013)
  [arXiv:1302.2807 [nucl-th]].  
  
\bibitem{Mart:2013gfa} 
  T.~Mart and A.~Sulaksono,
  %``Nonidentical protons,''
  Phys.\ Rev.\ C {\bf 87}, 025807 (2013)
  [arXiv:1302.6012 [nucl-th]].  
  
\bibitem{Mohr:2013axa} 
  P.~J.~Mohr, J.~Griffith and J.~Sapirstein,
  %``Bound-state field theory approach to proton structure effects in muonic hydrogen,''
  Phys.\ Rev.\ A {\bf 87}, no. 5, 052511 (2013)
  [arXiv:1304.2076 [hep-ph]].
  
  
\bibitem{Robson:2013nwa} 
  D.~Robson,
  %``Solution to the Proton Radius Problem,''
  arXiv:1305.4552 [nucl-th].  
  
  
  
\bibitem{Alarcon:2013cba} 
  J.~M.~Alarcon, V.~Lensky and V.~Pascalutsa,
  %``Chiral perturbation theory of muonic hydrogen Lamb shift: polarizability contribution,''
  Eur.\ Phys.\ J.\ C {\bf 74}, 2852 (2014)
  [arXiv:1312.1219 [hep-ph]].  
  
\bibitem{Downie:2013fya} 
  E.~J.~Downie, W.~J.~Briscoe, R.~Gilman and G.~Ron,
  %``Comment on ÒNonidentical protonsÓ,''
  Phys.\ Rev.\ C {\bf 88}, no. 5, 059801 (2013).  
   
  
  
\bibitem{Jentschura:2014ila} 
  U.~D.~Jentschura,
  %``Light Sea Fermions in Electron-Proton and Muon-Proton Interactions,''
  Phys.\ Rev.\ A {\bf 88}, 062514 (2013)
  [arXiv:1401.3666 [physics.atom-ph]].  
  
  
\bibitem{Eides:2014swa} 
  M.~I.~Eides,
  %``On Some Recent Ideas on the Proton Radius Puzzle and Lepton Anomalous Magnetic Moments,''
  arXiv:1402.5860 [hep-ph].  
  
\bibitem{Peset:2014yha} 
  C.~Peset and A.~Pineda,
  %``Model independent determination of the muonic hydrogen Lamb shift and proton radius,''
  arXiv:1403.3408 [hep-ph].  
  
\bibitem{Gainutdinov:2014kma} 
  R.~K.~.Gainutdinov and A.~A.~Mutygullina,
  %``Precision laser spectroscopy and the proton radius puzzle,''
  Bull.\ Russ.\ Acad.\ Sci.\ Phys.\  {\bf 78}, 189 (2014)
  [Izv.\ Ross.\ Akad.\ Nauk Ser.\ Fiz.\  {\bf 78}, 289Ð292 (2014)].
  
\bibitem{Tomalak:2014dja} 
  O.~Tomalak and M.~Vanderhaeghen,
  %``Two-photon exchange corrections in elastic muon-proton scattering,''
  Phys.\ Rev.\ D {\bf 90}, 013006 (2014)
  [arXiv:1405.1600 [hep-ph]].  
  
\bibitem{Pachucki:2014zea} 
  K.~Pachucki and K.~A.~Meissner,
  %``Proton charge radius and the perturbative quantum electrodynamics,''
  arXiv:1405.6582 [hep-ph].  
  
\bibitem{Glazek:2014ria} 
  S.~D.~Glazek,
  %``Calculation of size for bound-state constituents,''
  arXiv:1406.0127 [hep-th].  
  
\bibitem{Gorchtein:2014hla} 
  M.~Gorchtein,
  %``Forward sum rule for the $2\gamma$-exchange correction to the charge radius extraction from elastic electron scattering,''
  arXiv:1406.1612 [nucl-th].  
   
\bibitem{Peset:2014jxa} 
  C.~Peset and A.~Pineda,
  %``The two-photon exchange contribution to muonic hydrogen from chiral perturbation theory,''
  arXiv:1406.4524 [hep-ph].   
   

%%%%%%%%%%%%%%%%%%%%%%%%%%%  
  
 \bibitem{Jaeckel:2010xx} 
  J.~Jaeckel and S.~Roy,
  %``Spectroscopy as a test of Coulomb's law: A Probe of the hidden sector,''
  Phys.\ Rev.\ D {\bf 82}, 125020 (2010)
  [arXiv:1008.3536 [hep-ph]]. 
  
\bibitem{Kauffmann:2010cu} 
  S.~K.~Kauffmann,
  %``Do experiment and the correspondence principle oblige revision of relativistic quantum theory?,''
  Prespace.\ J.\  {\bf 1}, 1295 (2010)
  [arXiv:1009.3584 [physics.gen-ph]].  
  
\bibitem{Brax:2010gp} 
  P.~Brax and C.~Burrage,
  %``Atomic Precision Tests and Light Scalar Couplings,''
  Phys.\ Rev.\ D {\bf 83}, 035020 (2011)
  [arXiv:1010.5108 [hep-ph]].
  %%CITATION = ARXIV:1010.5108;%%  
  
\bibitem{Barger:2010aj} 
  V.~Barger, C.~-W.~Chiang, W.~-Y.~Keung and D.~Marfatia,
  %``Proton size anomaly,''
  Phys.\ Rev.\ Lett.\  {\bf 106}, 153001 (2011)
  [arXiv:1011.3519 [hep-ph]].  
  
\bibitem{TuckerSmith:2010ra} 
  D.~Tucker-Smith and I.~Yavin,
  %``Muonic hydrogen and MeV forces,''
  Phys.\ Rev.\ D {\bf 83}, 101702 (2011)
  [arXiv:1011.4922 [hep-ph]].  
  
\bibitem{Batell:2011qq} 
  B.~Batell, D.~McKeen and M.~Pospelov,
  %``New Parity-Violating Muonic Forces and the Proton Charge Radius,''
  Phys.\ Rev.\ Lett.\  {\bf 107}, 011803 (2011)
  [arXiv:1103.0721 [hep-ph]].
  
\bibitem{Rivas:2011dm} 
  J.~I.~Rivas, A.~Camacho and E.~Goeklue,
  %``Quantum spacetime fluctuations: Lamb Shift and hyperfine structure of the hydrogen atom,''
  Phys.\ Rev.\ D {\bf 84}, 055024 (2011)
  [arXiv:1105.6345 [gr-qc]].  
  
\bibitem{Barger:2011mt} 
  V.~Barger, C.~-W.~Chiang, W.~-Y.~Keung and D.~Marfatia,
  %``Constraint on parity-violating muonic forces,''
  Phys.\ Rev.\ Lett.\  {\bf 108}, 081802 (2012)
  [arXiv:1109.6652 [hep-ph]].  
  
\bibitem{Carlson:2012pc} 
  C.~E.~Carlson and B.~C.~Rislow,
  %``New Physics and the Proton Radius Problem,''
  Phys.\ Rev.\ D {\bf 86}, 035013 (2012)
  [arXiv:1206.3587 [hep-ph]].  
  
\bibitem{Wang:2013fma} 
  L.~-B.~Wang and W.~-T.~Ni,
  %``Proton radius puzzle and large extra dimensions,''
  Mod.\ Phys.\ Lett.\ A {\bf 28}, 1350094 (2013)
  [arXiv:1303.4885 [hep-ph]].  
  
\bibitem{Li:2013dwa} 
  Z.~Li and X.~Chen,
  %``Can Large Extra Dimensions Solve the Proton Radius Puzzle?,''
  arXiv:1303.5146 [hep-ph].  
  
\bibitem{Moumni:2013yta} 
  M.~Moumni and A.~BenSlama,
  %``Effects of Non-Commutativity on Light-Hydrogen-Like Atoms and Proton Radius,''
  Int.\ J.\ Mod.\ Phys.\ A {\bf 28}, 1350139 (2013)
  [arXiv:1305.3508 [hep-ph]].  
  
\bibitem{Carlson:2013mya} 
  C.~E.~Carlson and B.~C.~Rislow,
  %``Constraints to new physics models for the proton charge radius puzzle from the decay $K^+ \rightarrow \mu^+ +\nu + e^- + e^+$,''
  Phys.\ Rev.\ D {\bf 89}, 035003 (2014)
  [arXiv:1310.2786 [hep-ph]].  
  
\bibitem{Chang:2013yva} 
  W.~-F.~Chang, J.~N.~Ng and J.~M.~S.~Wu,
  %``Some consequences of the Majoron being the dark radiation,''
  Phys.\ Lett.\ B {\bf 730}, 347 (2014)
  [arXiv:1310.6513 [hep-ph]].  
  
\bibitem{Onofrio:2013fea} 
  R.~Onofrio,
  %``Proton radius puzzle and quantum gravity at the Fermi scale,''
  Europhys.\ Lett.\  {\bf 104}, 20002 (2013)
  [arXiv:1312.3469 [hep-ph]].  
  
\bibitem{Karshenboim:2014tka} 
  S.~G.~Karshenboim, D.~McKeen and M.~Pospelov,
  %``Constraints on muon-specific dark forces,''
  arXiv:1401.6154 [hep-ph].
  
\bibitem{Ubachs:2013kpa} 
  W.~Ubachs, W.~Vassen, E.~J.~S.~and and K.~S.~E.~Eikema,
  %``Precision metrology on the hydrogen atom in search for new physics,''
  Annalen Phys.\  {\bf 525}, A113 (2013).  
  
 \bibitem{Brax:2014zba} 
  P.~Brax and C.~Burrage,
  %``Explaining the Proton Radius Puzzle with Disformal Scalars,''
  arXiv:1407.2376 [hep-ph].   
  
  
  

\bibitem{Hill:2010yb} 
  R.~J.~Hill and G.~Paz,
  %``Model independent extraction of the proton charge radius from electron scattering,''
  Phys.\ Rev.\ D {\bf 82}, 113005 (2010)
  [arXiv:1008.4619 [hep-ph]].
  %%CITATION = ARXIV:1008.4619;%%
  
\bibitem{Arrington:2007ux}
  J.~Arrington, W.~Melnitchouk and J.~A.~Tjon,
  %``Global analysis of proton elastic form factor data with two-photon exchange
  %corrections,''
  Phys.\ Rev.\  C {\bf 76}, 035205 (2007)
  [arXiv:0707.1861 [nucl-ex]].
  %%CITATION = PHRVA,C76,035205;%%
  
\bibitem{Lung:1992bu} 
  A.~Lung, L.~M.~Stuart, P.~E.~Bosted, L.~Andivahis, J.~Alster, R.~G.~Arnold, C.~C.~Chang and F.~S.~Dietrich {\it et al.},
  %``Measurements of the electric and magnetic form-factors of the neutron from Q**2 = 1.75-GeV/c**2 to 4-GeV/c**2,''
  Phys.\ Rev.\ Lett.\  {\bf 70}, 718 (1993).
  %%CITATION = PRLTA,70,718;%%    
  
\bibitem{Gao:1994ud} 
  H.~Gao, J.~Arrington, E.~J.~Beise, B.~Bray, R.~W.~Carr, B.~W.~Filippone, A.~Lung and R.~D.~McKeown {\it et al.},
  %``Measurement of the neutron magnetic form-factor from inclusive quasielastic scattering of polarized electrons from polarized He-3,''
  Phys.\ Rev.\ C {\bf 50}, 546 (1994).
  %%CITATION = PHRVA,C50,546;%% 
  
 \bibitem{Anklin:1994ae} 
  H.~Anklin, D.~Fritschi, J.~Jourdan, M.~Loppacher, G.~Masson, I.~Sick, E.~E.~W.~Bruins and F.~C.~P.~Joosse {\it et al.},
  %``Precision measurement of the neutron magnetic form-factor,''
  Phys.\ Lett.\ B {\bf 336}, 313 (1994).
  %%CITATION = PHLTA,B336,313;%% 
  
\bibitem{Anklin:1998ae} 
  H.~Anklin, L.~J.~deBever, K.~I.~Blomqvist, W.~U.~Boeglin, R.~Bohm, M.~Distler, R.~Edelhoff and J.~Friedrich {\it et al.},
  %``Precise measurements of the neutron magnetic form-factor,''
  Phys.\ Lett.\ B {\bf 428}, 248 (1998).
  %%CITATION = PHLTA,B428,248;%%  


 \bibitem{Kubon:2001rj} 
  G.~Kubon, H.~Anklin, P.~Bartsch, D.~Baumann, W.~U.~Boeglin, K.~Bohinc, R.~Bohm and M.~O.~Distler {\it et al.},
  %``Precise neutron magnetic form-factors,''
  Phys.\ Lett.\ B {\bf 524}, 26 (2002)
  [nucl-ex/0107016].
  %%CITATION = NUCL-EX/0107016;%% 

\bibitem{Anderson:2006jp} 
  B.~Anderson {\it et al.}  [Jefferson Lab E95-001 Collaboration],
  %``Extraction of the neutron magnetic form-factor from quasi-elastic polarized-He-3(polarized-e, e-prime) at Q**2 = 0.1 - 0.6 (GeV/c)^2,''
  Phys.\ Rev.\ C {\bf 75}, 034003 (2007)
  [nucl-ex/0605006].
  %%CITATION = NUCL-EX/0605006;%%
  
  
\bibitem{Lachniet:2008qf} 
  J.~Lachniet {\it et al.}  [CLAS Collaboration],
  %``A Precise Measurement of the Neutron Magnetic Form Factor G**n(M)in the Few-GeV**2 Region,''
  Phys.\ Rev.\ Lett.\  {\bf 102}, 192001 (2009)
  [arXiv:0811.1716 [nucl-ex]].
  %%CITATION = ARXIV:0811.1716;%%
  
  
  
\bibitem{Bernauer:2010wm} 
  J.~C.~Bernauer {\it et al.}  [A1 Collaboration],
  %``High-precision determination of the electric and magnetic form factors of the proton,''
  Phys.\ Rev.\ Lett.\  {\bf 105}, 242001 (2010)
  [arXiv:1007.5076 [nucl-ex]].
  %%CITATION = ARXIV:1007.5076;%%  
  
 \bibitem{Borisyuk:2009mg}
  D.~Borisyuk,
  %``Proton charge and magnetic rms radii from the elastic $ep$ scattering
  %data,''
  Nucl.\ Phys.\  A {\bf 843}, 59 (2010)
  [arXiv:0911.4091 [hep-ph]].
  %%CITATION = NUPHA,A843,59;%% 
  
\bibitem{Belushkin:2006qa}
  M.~A.~Belushkin, H.~W.~Hammer and U.~G.~Meissner,
  %``Dispersion analysis of the nucleon form factors including meson continua,''
  Phys.\ Rev.\  C {\bf 75}, 035202 (2007)
  [arXiv:hep-ph/0608337].
  %%CITATION = PHRVA,C75,035202;%%
  
  \bibitem{Beringer:1900zz} 
  J.~Beringer {\it et al.}  [Particle Data Group Collaboration],
  %``Review of Particle Physics (RPP),''
  Phys.\ Rev.\ D {\bf 86}, 010001 (2012).
  %%CITATION = PHRVA,D86,010001;%%  
  
  
 \bibitem{Foldy:1952}
  L.~L.~Foldy,
  %``The Electromagnetic Properties  of Dirac Particles,''
  Phys.\ Rev.\  {\bf 87}, 688 (1952).
  %%CITATION = PHRVA,87,688;%%

\bibitem{Salzman:1955zz}
  G.~Salzman,
  %``Neutron-Electron Interaction,''
  Phys.\ Rev.\  {\bf 99}, 973 (1955).
  %%CITATION = PHRVA,99,973;%% 
  
\bibitem{Ernst:1960zza}
  F.~J.~Ernst, R.~G.~Sachs and K.~C.~Wali,
  %``Electromagnetic form factors of the nucleon,''
  Phys.\ Rev.\  {\bf 119}, 1105 (1960).
  %%CITATION = PHRVA,119,1105;%%  
  
  

  
  \bibitem{Hohler:1974ht}
  G.~Hohler and E.~Pietarinen,
  %``The Rho N N Vertex In Vector Dominance Models,''
  Nucl.\ Phys.\  B {\bf 95}, 210 (1975).
  %%CITATION = NUPHA,B95,210;%%

 \bibitem{Schwinger}
  J.~Schwinger,
  %``Field theory of unstable particles,''
  Annals Phys.\  {\bf 9}, 169-193 (1960).
  
\bibitem{Bhattacharya:2011ah} 
  B.~Bhattacharya, R.~J.~Hill and G.~Paz,
  %``Model independent determination of the axial mass parameter in quasielastic neutrino-nucleon scattering,''
  Phys.\ Rev.\ D {\bf 84}, 073006 (2011)
  [arXiv:1108.0423 [hep-ph]].
  %%CITATION = ARXIV:1108.0423;%%
  
  \bibitem{Federbush:1958zz}
  P.~Federbush, M.~L.~Goldberger and S.~B.~Treiman,
  %``Electromagnetic Structure of the Nucleon,''
  Phys.\ Rev.\  {\bf 112}, 642 (1958).
  %%CITATION = PHRVA,112,642;%%

\bibitem{Frazer:1960zzb}
  W.~R.~Frazer and J.~R.~Fulco,
  %``Effect of a Pion-Pion Scattering Resonance on Nucleon Structure. II,''
  Phys.\ Rev.\  {\bf 117}, 1609 (1960).
  %%CITATION = PHRVA,117,1609;%%
  
 \bibitem{Hohler}
G.~H\"{o}hler, Pion-nucleon scattering, 
in: H. Schopper (editor), Landolt-B\"ornstein database, 
Volume 9, subvolume b, part 1, Springer-Verlag, Berlin, 1983. 
[http://www.springermaterials.com/navigation/] 

\bibitem{Amendolia:1983di}
  S.~R.~Amendolia {\it et al.},
  %``Measurement Of The Pion Form-Factor In The Timelike Region For Q**2 Values
  %Between .1-Gev/C**2 And .18-Gev/C**2,''
  Phys.\ Lett.\  B {\bf 138}, 454 (1984).
  %%CITATION = PHLTA,B138,454;%%

\bibitem{Achasov:2005rg}
  M.~N.~Achasov {\it et al.},
  %``Study of the process e+ e- --> pi+ pi- in the energy region 400-MeV <
  %s**(1/2) < 1000-MeV,''
  J.\ Exp.\ Theor.\ Phys.\  {\bf 101}, 1053 (2005)
  [Zh.\ Eksp.\ Teor.\ Fiz.\  {\bf 101}, 1201 (2005)]
  [arXiv:hep-ex/0506076].
  %%CITATION = ZETFA,101,1201;%%

\bibitem{Cabibbo:1961sz}
 N.~Cabibbo and R.~Gatto,
 %``Electron Positron Colliding Beam Experiments,''
 Phys.\ Rev.\  {\bf 124}, 1577 (1961).
 %%CITATION = PHRVA,124,1577;%%
 
 \bibitem{Ablikim:2005nn}
  M.~Ablikim {\it et al.}  [BES Collaboration],
  %``Measurement of the cross section for e+ e- --> p anti-p at  center-of-mass
  %energies from 2.0-GeV to 3.07-GeV,''
  Phys.\ Lett.\  B {\bf 630}, 14 (2005)
  [arXiv:hep-ex/0506059].
  %%CITATION = PHLTA,B630,14;%%

\bibitem{Antonelli:1998fv}
  A.~Antonelli {\it et al.},
  %``The first measurement of the neutron electromagnetic form factors in  the
  %timelike region,''
  Nucl.\ Phys.\  B {\bf 517}, 3 (1998).
  %%CITATION = NUPHA,B517,3;%%
  
 
  
  
 \bibitem{Xu:2000xw} 
  W.~Xu, D.~Dutta, F.~Xiong, B.~Anderson, L.~Auberbach, T.~Averett, W.~Bertozzi and T.~Black {\it et al.},
  %``The Transverse asymmetry A(T-prime) from quasielastic polarized He-3 (polarized e, e-prime) process and the neutron magnetic form-factor,''
  Phys.\ Rev.\ Lett.\  {\bf 85}, 2900 (2000)
  [nucl-ex/0008003].
  %%CITATION = NUCL-EX/0008003;%%  
  
 \bibitem{Xu:2002xc} 
  W.~Xu {\it et al.}  [Jefferson Lab E95-001 Collaboration],
  %``PWIA extraction of the neutron magnetic form-factor from quasielastic polarized-He-3(polarized-e, e-prime) at Q**2 = 0.3-(GeV/c)**2 - 0.6-(GeV/c)**2,''
  Phys.\ Rev.\ C {\bf 67}, 012201 (2003)
  [nucl-ex/0208007].
  %%CITATION = NUCL-EX/0208007;%% 
 

\bibitem{CLAS}
\url{http://clasweb.jlab.org/cgi-bin/clasdb/msm.cgi?eid=111\&mid=1\&data=on}

\bibitem{Markowitz:1993hx} 
  P.~Markowitz, J.~M.~Finn, B.~D.~Anderson, H.~Arenhovel, A.~R.~Baldwin, D.~Barkhuff, K.~B.~Beard and W.~Bertozzi {\it et al.},
  %``Measurement of the magnetic form-factor of the neutron,''
  Phys.\ Rev.\ C {\bf 48}, 5 (1993).
  %%CITATION = PHRVA,C48,5;%%  
  



 \bibitem{Bruins:1995ns} 
  E.~E.~W.~Bruins, T.~S.~Bauer, H.~W.~den Bok, C.~P.~Duif, W.~C.~van Hoek, D.~J.~J.~de Lange, A.~Misiejuk and Z.~Papandreou {\it et al.},
  %``Measurement of the neutron magnetic form-factor,''
  Phys.\ Rev.\ Lett.\  {\bf 75}, 21 (1995).
  %%CITATION = PRLTA,75,21;%% 
  
\bibitem{Mcallister:1956ng} 
  R.~W.~Mcallister and R.~Hofstadter,
  %``Elastic Scattering of 188-{MeV} Electrons From the Proton and the $\alpha$ Particle,''
  Phys.\ Rev.\  {\bf 102}, 851 (1956).
  %%CITATION = PHRVA,102,851;%%  
  
  \bibitem{Aoki:2013ldr} 
  S.~Aoki, Y.~Aoki, C.~Bernard, T.~Blum, G.~Colangelo, M.~Della Morte, S.~DŸrr and A.~X.~El Khadra {\it et al.},
  %``Review of lattice results concerning low energy particle physics,''
  arXiv:1310.8555 [hep-lat].
  %%CITATION = ARXIV:1310.8555;%%
  
 \bibitem{Becher:2005bg} 
  T.~Becher and R.~J.~Hill,
  %``Comment on form-factor shape and extraction of |V(ub)| from B ---> pi l nu,''
  Phys.\ Lett.\ B {\bf 633}, 61 (2006)
  [hep-ph/0509090].
  %%CITATION = HEP-PH/0509090;%% 
  
 \bibitem{Sick:2005az} 
  I.~Sick,
  %``Nucleon radii,''
  Prog.\ Part.\ Nucl.\ Phys.\  {\bf 55}, 440 (2005).
  %%CITATION = PPNPD,55,440;%%  
  
\bibitem{Zhan:2011ji} 
  X.~Zhan, K.~Allada, D.~S.~Armstrong, J.~Arrington, W.~Bertozzi, W.~Boeglin, J.~-P.~Chen and K.~Chirapatpimol {\it et al.},
  %``High Precision Measurement of the Proton Elastic Form Factor Ratio $\mu_pG_E/G_M$ at low $Q^2$,''
  Phys.\ Lett.\ B {\bf 705}, 59 (2011)
  [arXiv:1102.0318 [nucl-ex]].
  %%CITATION = ARXIV:1102.0318;%%  
  
\bibitem{Lorenz:2012tm} 
  I.~T.~Lorenz, H.~-W.~Hammer and U.~-G.~Meissner,
  %``The size of the proton - closing in on the radius puzzle,''
  Eur.\ Phys.\ J.\ A {\bf 48}, 151 (2012)
  [arXiv:1205.6628 [hep-ph]].
  %%CITATION = ARXIV:1205.6628;%%  
  
 \bibitem{Beg:1964nm} 
  M.~A.~B.~Beg, B.~W.~Lee and A.~Pais,
  %``SU(6) and electromagnetic interactions,''
  Phys.\ Rev.\ Lett.\  {\bf 13}, 514 (1964).
  %%CITATION = PRLTA,13,514;%%
  
  \bibitem{Becchi:1965zza} 
  C.~Becchi and G.~Morpurgo,
  %``Test of the Nonrelativistic Quark Model for 'Elementary' Particles: Radiative Decays of Vector Mesons,''
  Phys.\ Rev.\  {\bf 140}, B687 (1965).
  %%CITATION = PHRVA,140,B687;%% 
  


\end{thebibliography}
\end{document}